\pdfoutput=1
\documentclass{emulateapj}
\usepackage{apjfonts}
\usepackage{amsmath,graphicx}
\newcommand{\beq}{\begin{equation}}
\newcommand{\eeq}{\end{equation}}
\newcommand{\bea}{\begin{eqnarray}}
\newcommand{\eea}{\end{eqnarray}}

\begin{document}

\title{Accretion onto ``Seed'' Black Holes in the First Galaxies}
\author{
Milo\v s Milosavljevi\'c\altaffilmark{1}, 
Volker Bromm\altaffilmark{1},
Sean M. Couch\altaffilmark{1}, and
S. Peng Oh\altaffilmark{2}
}
\altaffiltext{1}{Department of Astronomy, University of Texas, 1 University Station C1400, Austin, TX 78712.}
\altaffiltext{2}{Department of Physics, University of California, Santa Barbara, CA 93106.}
\righthead{ACCRETION ONTO SEED BLACK HOLES}
\lefthead{MILOSAVLJEVI\'C, ET AL.}

\begin{abstract}

The validity of the hypothesis that the massive black holes in high redshift quasars grew from stellar-sized ``seeds'' is contingent on a seed's ability to double its mass every few ten million years.  This requires that the seed accrete at approximately the Eddington-limited rate.  In the specific case of radiatively efficient quasiradial accretion in a metal-poor protogalactic medium, for which the Bondi accretion rate is often prescribed in cosmological simulations of massive black hole formation, we examine the effects of the radiation emitted near the black hole's event horizon on the structure of the surrounding gas flow. We find that photoheating and radiation pressure from photoionization significantly reduce the steady-state accretion rate and potentially render the quasiradial accretion flow unsteady and inefficient. The time-averaged accretion rate is always a small fraction of the ``Bondi'' accretion rate calculated ignoring radiative feedback.  The pressure of Ly$\alpha$ photons trapped near the \ion{H}{2} region surrounding the black hole may further attenuate the inflow.  These results suggest that an alternative to quasiradial, radiatively efficient Bondi-like accretion should be sought to explain the rapid growth of quasar-progenitor seed black holes.


\end{abstract}

\section{Introduction}
\label{sec:intro}
\setcounter{footnote}{0}

The origin and the early growth of massive black holes remains poorly understood.  The massive black holes in quasars ($M_{\rm BH}\sim 10^8-10^{10}M_\odot$), active galactic nuclei (AGN; $M_{\rm BH}>10^5M_\odot$), and the quiescent nuclei of nearby galaxies may have started out as stellar-mass ($M_{\rm BH}< 100\ M_\odot$) ``seed'' black holes \citep[e.g.,][]{Madau:01,Menou:01,Islam:03}. Is this plausible, that is, could the seed black holes have grown rapidly enough in the cosmic time available to them \citep[e.g.,][see also \citealt{Haiman:04}, \citealt{Djorgovski:08}, and references therein]{Haiman:01,Volonteri:03,Tanaka:08}? The rate at which a seed massive black hole can accrete is limited by the local density and the thermal structure of the protogalactic medium and by the effects of the radiation emitted near the event horizon on the accretion flow. Cosmological hydrodynamic simulations suggest that gravitational collapse produces dense central gas concentrations in protogalaxies \citep[e.g.,][]{Springel:05,Li:07,Pelupessy:07,Wise:07a,Wise:07b,Wise:08,DiMatteo:08,Greif:08}. Atomic densities have been found to reach $n\sim 10^4\textrm{ cm}^{-3}$ \citep[e.g.,][]{Greif:08}, and as much as $n\sim 10^6\textrm{ cm}^{-3}$ averaged over the central parsec around the potential minimum \citep{Bromm:03,Wise:07b,Wise:08}.  On spatial scales that are resolved in the simulations, gas is sufficiently concentrated to enable rapid accretion onto a seed black hole. An exception are the first hundred million years after the seed was formed, during which the surrounding gas density is lowered by the radiative feedback from the black hole's progenitor star (see, e.g., \citealt{Johnson:07,Alvarez:06,Alvarez:08}).  

Given an ample gas supply, will rapid accretion be inhibited by radiative effects? A reassessment of an accreting black hole's ability to control its own gas supply is needed to improve the realism of the treatment of black hole accretion in cosmological simulations. Existing cosmological simulations modeling the growth of seed black holes do not resolve the spatial scales on which some of the radiative processes may alter the accretion. The simulations also do not resolve the fine structure of the dense, turbulent, and possibly multiphase protogalactic medium in which the black holes are embedded. Semianalytic prescriptions are normally adopted for the accretion rate, but these prescriptions normally do not take into account the radiative feedback; it is normally assumed that a black hole, residing in a pressure-supported primordial gas cloud, can accrete steadily at the Bondi rate subject to the Eddington limit \citep[e.g.,][]{Volonteri:05,Alvarez:06,Johnson:07,Pelupessy:07,DiMatteo:08,Greif:08}. Here, we will evaluate the applicability of this assumption in view of the local radiative feedback that is present if the black hole accretes in a radiatively efficient fashion. We restrict our analysis to the early growth of quasar-progenitor ``seed'' black holes, which are occasionally referred to as ``miniquasars,'' where a stellar mass or an intermediate-mass black hole ($10^2M_\odot\lesssim M_{\rm BH}\lesssim 10^5M_\odot$) accretes from a metal and dust poor environment.  

The black hole's growth rate is particularly sensitive to the detailed thermal state of the irradiated accretion flow. This can be seen by noticing that the accretion rate, for quasiradial accretion, is influenced by the conditions at the sonic radius 
\beq
\label{eq:rs}
r_{\rm s}\sim 3\times10^{14}~ \frac{M_2}{T_{{\rm s},5}} \textrm{ cm} ,
\eeq
where $T_{\rm s} = 10^5 T_{{\rm s},5}\textrm{ K}$ is the temperature of the photoionized and photoheated flow at the sonic radius and $M_{\rm BH}=100\ M_2M_\odot$ is the black hole mass.  The sonic radius is normally unresolved in cosmological simulations of accretion onto black holes in the intermediate range of masses.\footnote{The recent, highly-resolved simulation of primordial protostar formation by \citet{Yoshida:08}, which has the requisite spatial resolution to resolve a sonic radius if it exists, only proceeds to the point where the initial hydrostatic core is formed, and does not treat the subsequent accretion flow onto the growing core.} It was recognized early that photoheating and photoionization pressure may prohibit steady radiatively efficient accretion \citep[e.g.,][]{Shvartsman:71,Buff:74,Hatchett:76,Ostriker:76}. It was suggested that accretion can still proceed at rates disallowed by the steady state solutions if cycles of rapid gas inflow, during pauses in accretion near the event horizon, alternate with photoheating or photoionization pressure-driven outflows \citep{Buff:74,Ostriker:76,Cowie:78,StellingwerfBuff:82,Begelman:85}. Such quasiperiodic cycling is seen in one-dimensional simulations of Compton-heated accretion onto $M_{\rm BH}\gtrsim 10^8M_\odot$ black holes in galaxy clusters, where the black hole accretes from a hot, ionized, and pressure supported atmosphere \citep{Ciotti:97,Ciotti:01,Ciotti:07,Sazonov:05}.   Recently, \citet{Ricotti:08} revisited the problem of irradiated quasiradial accretion in the context of primordial black hole growth following the cosmic recombination \citep[see, e.g.,][and references therein]{Ricotti:07}, and suggested that the accretion duty cycle is determined by the periodic formation of an \ion{H}{2} region surrounding the black hole.

It was further recognized that the formal existence of steady-state, spherically symmetric accretion solutions is sensitive to the treatment of boundary conditions far from the sonic radius \citep[e.g.,][]{Bisnovatyi:80}, and that these accretion flows can be locally thermally unstable \citep[e.g.,][]{Stellingwerf:82,Krolik:83} and should break down into time-dependent two-phase structure, containing a warm ionized phase and a hot, coronal phase \citep[e.g.,][]{Krolik:81}.  \citet{Wang:06} claimed that Compton heating in the vicinity of a seed massive black hole reduces the radial accretion rate to a small fraction of the Eddington-limited rate; we here suggest, however, that thermal runaway may engender the Compton-heated coronal phase only in metal-rich flows, where photoionization heating of the incompletely stripped oxygen drives gas heating beyond $\sim 10^5\textrm{ K}$ \citep[see, e.g.,][and \S~\ref{sec:photoionized_accretion} below]{Kallman:82}.

The impact of the radiation field produced near the event horizon on the accretion flow and on the state of the interstellar medium of the protogalaxy and that of the intergalactic medium \citep[e.g.,][]{Dijkstra:04,Kuhlen:05,Zaroubi:07,Thomas:08,Ripamonti:08,Spaans:08} is sensitive to the shape of the spectral energy distribution (SED) of the central source.  The SEDs of rapidly accreting low-mass massive black holes ($M_{\rm BH}\sim 10^5-10^6M_\odot$) exhibit significantly larger X-ray ($2\textrm{ keV}$) to optical spectral ratios than the AGN containing more massive rapidly accreting black holes \citep{Greene:07}, as is expected if a fraction of the radiation is produced in a geometrically thin disk. On the low mass end, if the microquasar SEDs \citep[e.g.,][and references therein]{Remillard:06} are an adequate prototype, the seed black hole SEDs may contain energetically significant components extending into the hard X-rays. The luminosity-weighted average spectrum of AGN containing intermediate mass black holes could differ substantially \citep[see, e.g.,][]{Venkatesan:01,Madau:04} from the scaled average quasar spectrum of \citet{Sazonov:04}. The ability of X-rays to escape the protogalaxy affects their contribution to the soft X-ray background \citep[e.g.,][]{Venkatesan:01,Dijkstra:04,Salvaterra:05} and the infrared background \citep{Cooray:04}. Another difference between the first AGN and the starburst or post-starburst AGN is related to the differences in metallicity of the accretion flows.  The thermal phase structure of the interstellar medium exposed to UV and X-ray radiation is sensitive to metal abundances, especially for $T<10^4\textrm{ K}$ \citep[e.g.,][]{Donahue:91} and for $T\gtrsim \textrm{ few }\times10^4\textrm{ K}$ \citep[e.g.,][and \S~\ref{sec:photoionized_accretion}]{Kallman:82}. 

The role of radiative feedback in the formation of the first massive black holes resembles the radiative regulation in the formation of the first massive protostars \citep{Omukai:01,Omukai:03a,Omukai:02,Omukai:03b,Tan:04,McKee:07}, though, of course, the central sources have very different spectra. In protostars, an \ion{H}{2} region forms around the protostar and the protostellar disk that feeds its growth; persistent accretion onto the protostar may be quenched by radiation pressure. The protostellar accretion is characterized by a lower radiative efficiency and a higher accretion rate than the black hole accretion for the same accretor mass. The growing protostar is embedded in a supersonically collapsing and very dense envelope from inception, whereas here we assume that the seed black hole is born in the collapse of a massive star without such an envelope.  The presence of an infalling envelope implies that the \ion{H}{2} region is initially \emph{trapped} near the protostar, inside the radius where the envelope infall velocity turns from subsonic to supersonic.  For accretion from a diffuse medium onto a seed black hole, the \ion{H}{2} region is inevitably extended and the flow crossing the ionization front is highly subsonic; this marks a crucial difference with the protostellar accretion scenarios. 
  
Ignoring radiative effects, accretion onto the black hole is quasiradial on certain length scales (e.g., $r_{\rm s}\lesssim r\lesssim 10\textrm{ pc}$) if turbulence in the gas is weak \citep{Krumholz:06} and if the gas is not rotationally supported on these scales.  If the accretion flow is shock-free (an unlikely condition) and possesses small net rotation, the buildup of vorticity near $r\sim r_{\rm s}$ may reduce the accretion rate by $\sim 60\%$ \citep{Krumholz:05}.  Quasiradial accretion may be expected even when the baryons in a protogalaxy initially form a rotating disk, because self-gravity in the gas on scales of the protogalactic disk destabilizes rotational equilibria to convert disk-like configurations into quasiradial, stratified, pressure supported, and possibly turbulent configurations.  The gas distribution could still be rotationally supported on larger scales, where, e.g., dark matter dominates gravity, and on much smaller scales, where the black hole dominates gravity.  This description may apply to high-redshift protogalaxies \citep[see, e.g.,][]{Oh:02,Volonteri:05,Wang:06}, and so here, we focus on angular momentum-free accretion and defer examining the role of angular momentum to a subsequent paper.  

The quasiradial gas flow can either be steady and radial, or unsteady and characterized by alternating inflow, outflow, and nonradial motions. In view of these possibilities, this work is organized as follows. In \S~\ref{sec:bondi} we attempt, and fail, to construct a steady, radial solution for accretion at high accretion rates and high radiative efficiencies.   In \S~\ref{sec:timedep} we provide a qualitative analysis of time-dependent, episodic accretion, and attempt to estimate the average accretion rate. In \S~\ref{sec:clumpy} we discuss the consequences of the presence of cold, inhomogeneous, and turbulent gas in the vicinity of the black hole.  In \S~\ref{sec:conclusions} summarize our main conclusions.  In Table \ref{tab:symbols} we present an overview of our notation.

\begin{deluxetable*}{lll}
\tablecolumns{3}
\tablecaption{Index of Notation\label{tab:symbols}}
\tablehead{\colhead{Quantity} & \colhead{Symbol} & \colhead{Note}}
\startdata

Acceleration due to gravity & $a_{\rm grav}$ & $-GM_{\rm BH}/r^2$ \\ 
Acceleration due to radiation pressure & $a_{\rm rad}$ & see text (\S~\ref{sec:prad_continuum}) \\
Spectral index of the SED & $\alpha$ & $F_\nu,\ f_\nu\propto \nu^{-\alpha}$ \\
Case $B$ recombination rate for hydrogen & $\alpha_B$ & $\approx 2.6\times10^{-13} T_4^{-1}\textrm{ cm}^3\textrm{ s}^{-1}$ \\
Collisional ionization rate for hydrogen & $\alpha_{\rm ion}$ & \nodata \\

Abundance of species $i$ relative to hydrogen & $\chi_i$ & $\equiv n_i/n_{\rm H}$ \\

Logarithmic slope of $T_{\rm s}$ as a function of $\epsilon$ & $\delta$ & $T_{\rm s}\propto \epsilon^\delta$ \\

Radiative efficiency & $\epsilon$ & $=L/\dot M c^2$ \\
Critical efficiency for radiation pressure suppression at low efficiencies & $\epsilon_{\rm crit}$ & $\phi r_{\rm ion}/r_{\rm s}=1$ \\
Ionization potential of species $i$ & $E_i$ & \nodata \\

Duty cycle for episodic accretion & $f_{\rm duty}$ & $\equiv\langle L^2\rangle/\langle L\rangle^2$  \\
Density enhancement in the ionized gas in episodic accretion & $f_{\rm epi}$ & $\geq 1$ \\
Fraction of photon energy going to photoionizations in the ionized gas & $f_{\rm ion}$ & $\sim 1/3$ \\
Fraction of photon energy reprocessed to Ly$\alpha$ & $f_{{\rm Ly}\alpha}$ & $\sim 2/3$ \\
Fraction of $L$ that reaches the edge of the \ion{H}{2} region & $f_{\rm res}$ & see text (\S~\ref{sec:prad_continuum}) \\
Pressure enhancement due to turbulence in the neutral gas & $f_{\rm turb}$ & $\sim 1+{\cal M}^2$ \\

Adiabatic index & $\gamma$ & $=5/3$ \\

Luminosity in units of the Eddington luminosity & $\ell$ & $L/L_{\rm Edd}$ \\
Luminosity of isothermal accretion ignoring radiative heating & $\ell_{\rm Bondi}$ & see text (\S~\ref{sec:rs_conditions}) \\
Ly$\alpha$-pressure-limited accretion rate & $\ell_{{\rm crit,Ly}\alpha}$ & see text (\S~\ref{sec:prad_lyalpha}) \\
Peak luminosity in episodic accretion & $\ell_{\rm max}$ & \nodata \\
Luminosity of steady-state photoheated accretion  & $\ell_{\rm s.s.}$ & see text (\S~\ref{sec:rs_conditions}) \\
Luminosity emitted at $r\lesssim r_{\rm disk}$ & $L$ & \nodata \\ 
Eddington luminosity for Thomson scattering & $L_{\rm Edd}$ & $=4\pi GM_{\rm BH} m_p c/\sigma_{\rm T}$ \\

Mean molecular mass of the ionized gas & $\mu$ & $\sim 0.6$ \\
Turbulent Mach number of the neutral gas & ${\cal M}$ & \nodata \\ 
Mass of the black hole & $M_{\rm BH}$ & \nodata \\
Central accretion rate & $\dot M$ & \nodata \\
Central accretion rate ignoring radiation & $\dot M_{\rm Bondi}$ & see text (\S~\ref{sec:rs_conditions}) \\

Density at the \ion{H}{2} region's edge in a protogalactic density cusp & $n_{\rm cusp}$ & see text (\S~\ref{sec:density}) \\
Density of the neutral gas and ambient density & $n_{\rm HI},\  n$ & \nodata \\
Particle density within the \ion{H}{2} region & $n_{\rm HII}$ & $f_{\rm turb}n_{\rm HI}T_{\rm HI}/T_{\rm HII}$ \\
Maximum density for high ionization at the sonic radius & $n_{\rm max,ion}$ & see text (\S~\ref{sec:rs_conditions}) \\
H$^-$-dissociating photon production rate & $\dot N_{\gamma,{\rm H}^-}$ & see text (\S~\ref{sec:photodissociation}) \\
Lyman-Werner photon production rate & $\dot N_{\rm LW}$ & see text (\S~\ref{sec:photodissociation}) \\
Total ionization rate in the \ion{H}{2} region & $\dot N_{\rm ion}$ & see text (\S~\ref{sec:hii_region}) \\
Total recombination rate in the \ion{H}{2} region & $\dot N_{\rm rec}$ & see text (\S~\ref{sec:hii_region}) \\
Number of Ly$\alpha$ reflections on \ion{H}{2} region's walls to escape & $N_{\rm reflect}$ & see text (\S~\ref{sec:prad_lyalpha}) \\

Dimensionless photoionization pressure acceleration in the \ion{H}{2} region & $\phi$ & see text (\S~\ref{sec:prad_continuum}) \\
Dimensionless radiation pressure acceleration at the \ion{H}{2} region's edge & $\psi$ & see text (\S~\ref{sec:prad_continuum}) \\
Pressure of the neutral gas & $P_{\rm gas}$ & $=n_{\rm HI} m_p k T_{\rm HI}$ \\
Pressure of the Ly$\alpha$ radiation & $P_{{\rm Ly}\alpha}$ & see text (\S~\ref{sec:prad_lyalpha}) \\

Radius of the \ion{H}{2} region in a protogalactic density cusp & $r_{\rm ion,cusp}$ & see text (\S~\ref{sec:density}) \\
Bondi radius ignoring radiative heating and acceleration & $r_{\rm B}$ & see text (\S~\ref{sec:duty_cycle}) \\
Sonic radius & $r_{\rm s}$ & $v(r_{\rm s})=c_{\rm s}(r_{\rm s})$ \\
Disk radius & $r_{\rm disk}$ & $\ll r_{\rm s}$ \\
Radius where thermal time equals inflow time & $r_{\rm equi}$ & $\max\{t_{\rm heat},t_{\rm cool}\}=r/v$ \\
Radius of the \ion{H}{2} region & $r_{\rm ion}$ & see text (\S~\ref{sec:hii_region}) \\
Width of the H$_2$ photodissociation shell & $\Delta r_{\rm diss}$ & see text (\S~\ref{sec:photodissociation}) \\
Width of the neutral shell surrounding the \ion{H}{2} region & $\Delta r_{\rm shell}$ & see text (\S~\ref{sec:prad_lyalpha}) \\

Line-center Ly$\alpha$ scattering cross section & $\sigma_0$ & $\approx 5.9\times10^{-14}T_4^{-1/2}$ \\
Flux-averaged photoionization cross section of species $i$ & $\bar \sigma_i$ & see text (\S~\ref{sec:prad_continuum}) \\
Photon number-averaged photoionization cross section  & $\tilde \sigma_i$ & see text (\S~\ref{sec:prad_continuum}) \\
H$_2$ formation suppression factor due to H$^-$ photodissociation & $S$ & see text (\S~\ref{sec:photodissociation}) \\

Ly$\alpha$ line-center optical depth of the neutral shell & $\tau_0$ & see text (\S~\ref{sec:prad_lyalpha}) \\
Bremsstrahlung cooling time & $t_{\rm Brems}$ & see text (\S~\ref{sec:rs_conditions}) \\
Compton heating time & $t_{\rm C}$ & see text (\S~\ref{sec:rs_conditions}) \\
Cooling time & $t_{\rm cool}$ & \nodata \\
Heating time & $t_{\rm heat}$ & \nodata \\
Photoionization heating time & $t_{\rm photo}$ & see text (\S~\ref{sec:inflow_outflow}) \\
Inflow time at the sonic radius & $t_{\rm s}$ & $\sim r_{\rm s}/c_{\rm s}(r_{\rm s})$ \\
Salpeter mass-exponentiation time scale & $t_{\rm Salp}$ & $=\epsilon M_{\rm BH}c^2/L_{\rm Edd}$ \\
Temperature of the ambient neutral gas & $T_{\rm HI}$ & \nodata \\
Temperature within the \ion{H}{2} region & $T_{\rm HII}$ & \nodata \\
Temperature at the sonic radius & $T_{\rm s}$ & see text (\S~\ref{sec:intro}) \\

Density enhancement over isothermal accretion with $T_{\rm HII}=T_{\rm s}$ & $\Upsilon$ & $\gtrsim 1$\\

Radial inflow velocity & $v$ & \nodata \\

Dimensionless inflow velocity & $w$ & $\equiv v/c_{\rm s}(r_{\rm s})$ \\

Ionization parameter & $\xi$ & $\equiv L/r^2n$ \\
Dimensionless ionization parameter & $\Xi$ & $\equiv \xi/4\pi kT_{\rm HII} c$ \\

Dimensionless radius & $y$ & $\sim r/r_{\rm s}$ \\

Metallicity & $Z$ & \nodata \\

\enddata
\end{deluxetable*}

\section{The Prospect of Time-Independent Accretion}
\label{sec:bondi}

In this section we set out to test the model, ubiquitous in semi-analytic and semi-numerical studies of massive black hole evolution, in which the accretion onto a black hole from the protogalactic medium is steady and quasiradial. In \S~\ref{sec:photoionized_accretion} we review the standard theory of photoionized radial accretion, and pay particular attention to the dependence of the structure of the flow on the metallicity of the accreting gas.  We find that at low metallicities, the flow evades thermal runaway and heating to the Compton temperature at least until it passes the sonic radius. In \S~\ref{sec:hii_region} we estimate the size of the \ion{H}{2} region surrounding the black hole and derive conditions under which the self-gravity within the ionized sphere can be ignored.  In \S~\ref{sec:rs_conditions}, we take a closer look at the state of the gas as it passes the sonic radius and check whether it is in local thermal and statistical equilibrium. In \S~\ref{sec:density} we justify our reference choice for the ambient density in the neutral medium surrounding the \ion{H}{2} region.  This justification is necessary because the central gravitational collapse in a protogalaxy will yield a wide range of densities, yet most of our estimates depend on a specific choice of density.  In \S~\ref{sec:prad_continuum} we study the effects of the photoionization radiation pressure within the \ion{H}{2} region on the structure of the steady-state accretion flow. We find that, depending on the parameters, photoionization radiation pressure in the outer parts of the \ion{H}{2} region may prevent accretion at near the Eddington-limited rate.  In \S~\ref{sec:radiatively_inefficient}, we briefly address the case of radiatively-inefficient accretion. In \S~\ref{sec:prad_lyalpha} we estimate the pressure of Ly$\alpha$ line radiation produced in and near the \ion{H}{2} region and confined by resonance line scattering in its vicinity. We find that Ly$\alpha$ radiation pressure can exceed the thermal gas pressure in the \ion{H}{2} region, and this presents an additional challenge to strictly stationary quasiradial solutions. 

\subsection{Photoionized Quasiradial Accretion}
\label{sec:photoionized_accretion}

We ignore the angular momentum of the gas, which is assumed to be free of metals and dust, on radial length scales $\gtrsim r_{\rm s}$ and assume that the accretion flow becomes rotationally supported and collapses into a hypothetical geometrically thin disk only at radii $\lesssim r_{\rm disk}\ll r_{\rm s}$.  Furthermore, we assume that the disk accretes onto the black hole with a high radiative efficiency $\epsilon$, as is expected for thin-disk accretion, such that the bolometric outward radiation flux passing through radius $r_{\rm s}$ is $F(r_{\rm s}) = \epsilon \dot M(r_{\rm s})c^2 /4\pi r_{\rm s}^2$. 
We assume that, absent feedback effects, the density scale-height is much larger than any other length scale under consideration, so that the ambient medium has effectively constant density. 
We restrict our attention to the accretion flow at radii $r> r_{\rm s}$, where we assume that the flow is quasiradial and exposed to the radiation emitted by the disk at $r\sim 0$.   Without angular momentum, the radial support against the black hole's gravity must arise from gas pressure gradients and the radiation pressure force. Conservation laws yield the relation \citep[see, e.g.,][]{Ostriker:76,Lamers:99}
\beq
\label{eq:inflow}
\frac{dv}{dr}\left(v^2-\frac{\gamma ~k~T}{\mu ~m_p}\right)
=  v\left(\frac{2~\gamma k~T}{r~\mu ~m_p} + a_{\rm tot}\right) + (\gamma - 1) (H-C) ,
\eeq
where $v$ is the radial inflow velocity, which is positive when the flow is directed inward, $\gamma$ the adiabatic index that defines the relation of internal energy density to pressure, $k$ the Boltzmann constant, $\mu$ the mean molecular mass, $m_p$ the proton mass, $T$ the gas temperature, $a_{\rm tot}=a_{\rm rad}+a_{\rm grav}$ the sum of the accelerations due to gravity and radiation pressure, $H$ the photoheating rate, and $C$ the cooling rate (both per unit mass).

As we show in \S~\ref{sec:hii_region}, 
there exists a radius $r_{\rm equi}$ outside of which the heating time $t_{\rm heat}\sim kT/(\gamma-1) \mu m_p H$ and cooling time $t_{\rm cool}\sim kT/(\gamma-1) \mu m_p C$ are much shorter than the inflow time $\sim r/v$.  Since photochemical time scales in dense, ionized gas are generally short, outside this radius the gas reaches an approximate local thermal and statistical equilibrium, and so the last term in equation (\ref{eq:inflow}) that is proportional to $H-C$ can be dropped. An estimate that we provide at the end of 
\S~\ref{sec:hii_region}
below suggests that for metal-poor accretion, the radius $r_{\rm equi}$ is usually at most slightly larger than the radius $r_{\rm s}$ at which the flow becomes supersonic.  We ignore this complication and assume $r_{\rm equi}\lesssim r_{\rm s}$.

The flow is not adiabatic, but we let $c_{\rm s}\equiv (\gamma kT/\mu m_p)^{1/2}$ denote the usual adiabatic sound speed.  Continuity at the sonic radius $r_{\rm s}$ where $v=c_{\rm s}$ requires that $2 [c_{\rm s}(r_{\rm s})]^2 r_{\rm s}^{-1}+a_{\rm grav}(r_{\rm s})+a_{\rm rad}(r_{\rm s})=0$, and because the self gravity of the gas is negligible, $a_{\rm grav}(r_{\rm s})=-GM_{\rm BH}/r_{\rm s}^2$.  Since the gas is almost fully ionized at $r\sim r_{\rm s}$, the radiation pressure at the sonic radius is mainly due to electron scattering.  Then $a_{\rm rad}(r_{\rm s})=- \ell(r_{\rm s}) a_{\rm grav}(r_{\rm s})$, where $\ell(r)\equiv L(r)/L_{\rm Edd}$ is the ratio of the total luminosity to the Eddington luminosity for all opacities. Since under a wide range of conditions \citep[see, e.g.,][]{Blondin:86}, the flow at the sonic radius is optically thin to electron scattering, provided that the small accretion disk inside the sonic radius does not shadow and reprocess to low frequencies a substantial fraction of the central luminosity, we can assume that $L(r_{\rm s})=L(0)$. Then, in a steady state, the luminosity can be related to the total mass flux into the central source, $L(0)=\epsilon \dot M c^2= 4\pi \epsilon r_{\rm s}^2 c_{\rm s}(r_{\rm s}) \mu n(r_{\rm s}) m_p c^2$, where $n(r)$ is the gas number density.

\begin{figure}
\begin{center}
\includegraphics[width=3.0in]{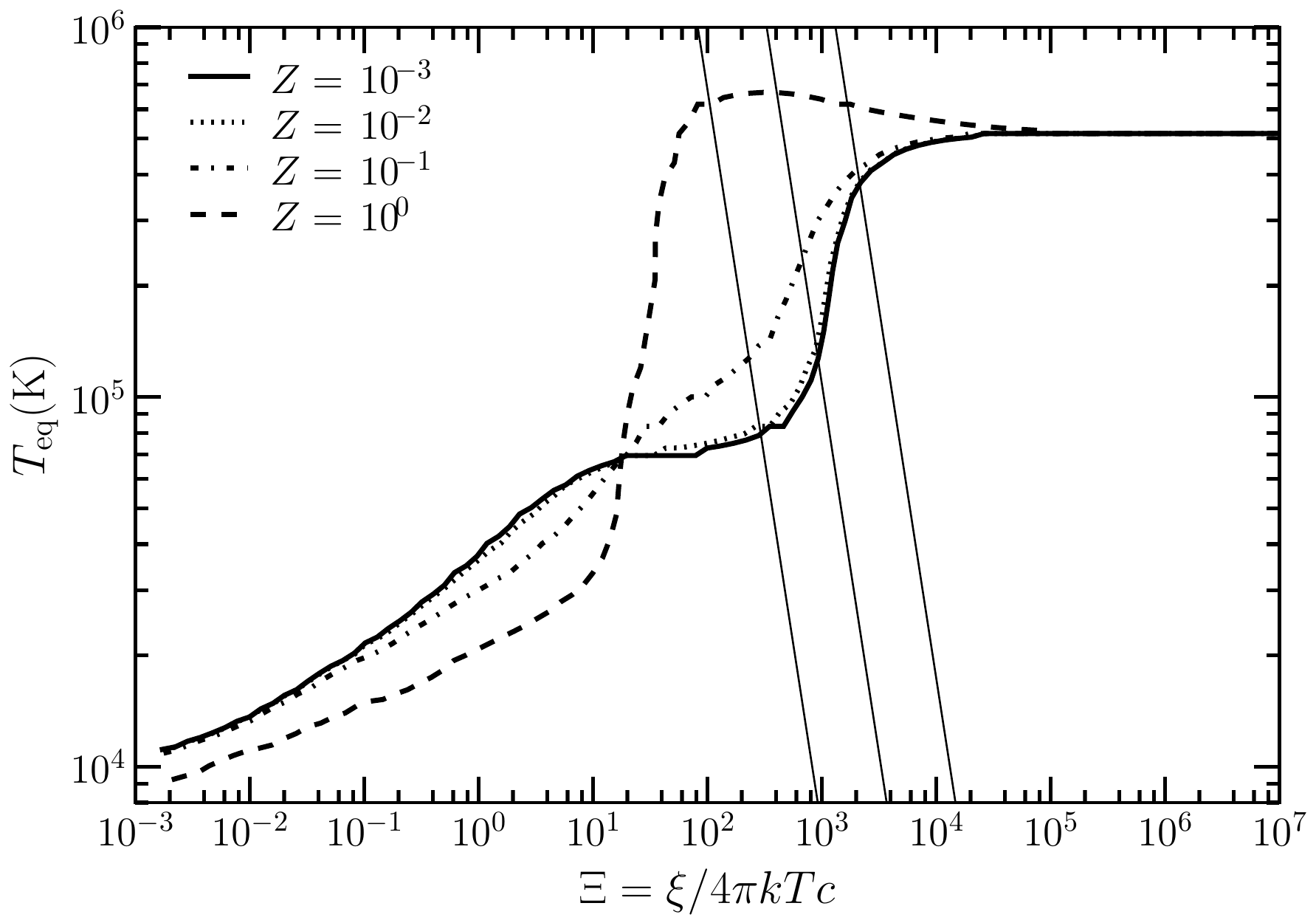}
\end{center}

\caption{Metallicity dependence of the local thermal and statistical equilibrium temperature of a photoionized gas under optically thin conditions as a function of the ionization parameter $\Xi$ (see text). The functions $T_{\rm eq}(\Xi)$ for four gas metallicities, expressed in units of the solar metallicity, were calculated with the photoionization code XSTAR \citep{Kallman:01} for an $f_\nu\propto \nu^{-1.5}$ spectrum between $0.1\textrm{ Ryd}$ and $1,000\textrm{ Ryd}$.  The curves $T_{\rm eq}(\Xi)$ are single valued and thus all equilibria are stable for the particular choice of spectrum, but they need not all be stable for harder spectra. The plot shows that $T_{\rm eq}(\Xi)$ is nearly independent of metallicity for $Z\lesssim 0.01\ Z_\odot$, and that a hot coronal phase appears at $\Xi \sim 10^3$ in a gas with $Z\lesssim 0.1\ Z_\odot$ and at $\Xi\sim 10$ in a more metal rich gas.  The straight lines are the relations $\Xi=\epsilon c(\gamma\mu m_p/kT)^{1/2}$ that must hold at the sonic radius for values of the radiative efficiency, from left to right, of $\epsilon=0.025$, $\epsilon=0.1$, and $\epsilon=0.4$. The temperature at the sonic radius is found at the intersection with the $T_{\rm eq}(\Xi)$ curve.}
\label{fig:tofxi}
\end{figure}

Let $\Xi\equiv L/4\pi r^2 n k T c$ denote the dimensionless ionization parameter introduced in \citet{Krolik:81}, which is the ratio of the radiative momentum flux to the gas pressure.  Then at the sonic radius $\Xi_{\rm s} =  \epsilon \gamma c / c_{\rm s}(r_{\rm s})$ (see straight lines in Fig.~\ref{fig:tofxi}).  In a chemical equilibrium determined purely by two-body collisional processes and photoionization, the equilibrium abundances $\chi_i \equiv n_i/n$ of all species are functions of the temperature and $F_\nu/n$ only, where $F_\nu$ is the radiation flux at frequency $\nu$ such that $\int F_\nu d\nu=F=L/4\pi r^2$. On the other hand, the temperature $T_{\rm eq}$ arising from the equilibrium of photoheating and two-body collisional cooling is a function of $\chi_i$ and $F_\nu/n$ only.  Therefore, for a particular SED $f_\nu = F_\nu/F$, the equilibrium temperature is determined only by $L/4\pi r^2 n$, i.e., $T_{\rm eq}$ lies in one-to-one relation with $\Xi$ and we can write $T_{\rm eq}=T_{\rm eq}(\Xi;f_\nu)$. For particularly hard spectra, instead of being one-to-one, the function $T_{\rm eq}(\Xi;f_\nu)$ can be multivalued in a certain range of $\Xi$.  Then, the temperature of a fluid element depends on its thermal history. Since thermodynamic perturbations at constant $\Xi$ are isobaric, the possibly multivalued function $T_{\rm eq}(\Xi)$ at fixed $f_\nu$ determines the thermal phase structure. 

For a fixed SED, $T_{\rm eq}(\Xi)$ is sensitive to the metallicity for $10^5\textrm{ K}\lesssim T\lesssim 10^6 \textrm{ K}$ because helium line cooling and Bremsstrahlung cooling dominate the cooling rate at any metallicity (with a comparable contribution from iron at high temperatures in the metal-enriched case), while oxygen and iron photoionization heating are by far the most important heating processes in the metal rich case \citep[e.g.,][]{Kallman:82}.   Figure \ref{fig:tofxi} shows that for the $f_\nu\propto \nu^{-1.5}$ spectrum for $0.1\textrm{ Ryd}< h\nu<10^3\textrm{ Ryd}$, at near-solar metallicities, the Compton-heated hot coronal phase appears at $\Xi\sim 10$ \citep[as is well known, see, e.g.,][]{Krolik:81}, whereas in a gas with subsolar metallicity $Z\lesssim 0.1\ Z_\odot$, the coronal phase does not appear until the ionization parameter reaches $\Xi\sim 10^3$.  The optically thin $T_{\rm eq}(\Xi)$ becomes independent of metallicity at $Z\sim 0.01\ Z_\odot$.

In Figure \ref{fig:tofxi}, we also show the relations $\Xi=\epsilon c(\gamma\mu m_p/kT)^{1/2}$ that must hold at the sonic radius for three values of the radiative efficiency: $\epsilon=0.025$, $\epsilon=0.1$, and $\epsilon=0.4$, which, speculatively, might be expected for a rapidly rotating black hole with a geometrically thin retrograde disk, a nonrotating black hole with a thin disk, and a rapidly rotating black hole with a geometrically thin prograde disk, respectively \citep[see, e.g.,][and references therein]{Novikov:73,Zhang:97,Beckwith:06,Beckwith:08,Noble:08}. The temperature at the sonic radius is found at the intersection with the $T_{\rm eq}(\Xi)$ curve, that is, we have the implicit relation $T_{\rm s}=T_{\rm eq}[\Xi_{\rm s};f_\nu(r_{\rm s})]=T_{\rm eq}[\epsilon c(\gamma \mu m_{\rm p}/kT_{\rm s})^{1/2}; f_\nu]$, which can be solved for the temperature at the sonic radius $T_{\rm s}$ as a function of $\epsilon$ and $f_\nu$ \citep[e.g.,][]{Ostriker:76}. Evidently, in metal poor gas, the 
equilibrium temperature at the sonic radius is a strong function of the radiative efficiency; highly radiatively efficient accretion is susceptible to thermal runaway, where the ionization state converges to full ionization as the Compton heating overtakes thermal evolution and the gas becomes fully ionized \citep[see, e.g.,][and references therein]{Krolik:99}. Note, however, that the gas may not attain the Compton temperature if the inflow time becomes shorter than the heating time for combined Compton and photoionization heating (see \S~\ref{sec:hii_region}).  Given $T_{\rm s}$ and the sonic radius determined from the relation involving $c_{\rm s}(r_{\rm s})$ and the forces acting on the gas, we have that $r_{\rm s}=(1-\ell)\mu m_p GM_{\rm BH}/2\gamma k T_{\rm s}$.  For example, for power law SEDs $F_\nu \propto \nu^{-1.5}$ and radiative efficiencies $\epsilon\sim 0.1$, typical temperatures at the sonic radius are $T_{\rm s}\sim 10^5 \textrm{ K}$, and so the sonic radii are $r_{\rm s}\sim3\times 10^{14} M_2 \textrm{ cm}$.

We can write the usual ionization parameter $\xi \equiv L/r^2 n=4\pi kTc\Xi$ of \citet{Tarter:69}\footnote{\citet{Tarter:69} define $\xi$ in terms of the hydrogen density $n_{\rm H}$ and we do so in terms of the total particle density $n$; in what follows, we do not always rigorously track the dependence on the molecular weight $\mu$.} as $\xi = 4\pi v \mu m_p L /\dot M$, which in the optically thin limit becomes $\xi = 4\pi v \mu m_p \epsilon c^2$.  In this limit, the temperature $T_{\rm eq}$, which is in one-to-one relation with $\xi$, is also in one-to-one relation with the velocity, $T_{\rm eq}(v;f_\nu,\epsilon)$, and so one can write equation (\ref{eq:inflow}) as a differential equation with a single unknown function $v(r)$.  When the flow is isothermal, we have the well-known asymptotic solution far from the black hole
\beq
\label{eq:velocity_asymptotic_iso}
v \sim e^{3/2} \left(\frac{r}{r_{\rm s}}\right)^{-2} c_{\rm s}(r_{\rm s}) \ \ \ \  (r\gg r_{\rm s}, \textrm{ isothermal}) .
\eeq
If the temperature of the photoionized flow can decrease with radius and thus the infalling gas can acquire momentum before it heats to $\sim T_{\rm s}$, the asymptotic velocity $v(r)$ far from the black hole can exceed the isothermal value given in equation (\ref{eq:velocity_asymptotic_iso}) by a factor of several.  We thus write 
\beq
\label{eq:velocity_asymptotic}
v \sim 4.5\ \Upsilon  \left(\frac{r}{r_{\rm s}}\right)^{-2}~ c_{\rm s}(r_{\rm s}) \ \ \ \  (r\gg r_{\rm s}) ,
\eeq
where $\Upsilon\gtrsim 1$.

Ideally, we would like to match this solution to the conditions far from the black hole, where the density and the total pressure are $n_\infty$ and $P_\infty$.  One could attempt to set the boundary conditions $n(r_1)=n_\infty$ and $n(r_1) k T_{\rm eq}(r_1)=P_\infty\equiv n_\infty k T_\infty $ at some radius $r_1$, where the last relation defines $T_\infty$.  This may indeed be possible at very low densities or very low radiative efficiencies.  At high densities and efficiencies, however, the photoionization equilibrium at some well defined radius $r_{\rm ion}\gg r_{\rm s}$ abruptly transitions into a neutral state, i.e., on its way toward the black hole the gas passes a quasistationary ionization front.  Gas density, velocity, temperature, and pressure may be discontinuous at the ionization front; furthermore, the warm or cold neutral gas in the immediate vicinity of the ionized region may be supersonically turbulent, in which case the gas density is inhomogeneous and the ram pressure of turbulent flows cannot be neglected.  We proceed to an attempt to determine under which conditions is a steady, strictly time-independent accretion across the stationary ionization front possible.

\subsection{Size of the \ion{H}{2} Region around the Black Hole}
\label{sec:hii_region}

We will assume that the ionized region is surrounded by warm, partially-ionized gas, though in reality, the accreting black hole and its ionization sphere may be embedded in a cold, molecular, and supersonically turbulent medium. However, since the UV radiation from the black hole and the \ion{He}{2} recombination radiation from the photoionization annulus (see \S~\ref{sec:photodissociation} below) will dissociate molecules in the vicinity of the \ion{H}{2} region, a layer of warm, atomic gas should surround the ionization sphere even if the molecular phase exists at somewhat larger optical depths.  We further assume that far from the sonic radius, where the black hole's gravity can be ignored, the ionized gas is in gas pressure equilibrium with the surroundings, i.e., 
\beq
T_{\rm HII}~n_{\rm HII}=f_{\rm turb}~T_{\rm HI}~n_{\rm HI} \ \ \ \  (r\gg r_{\rm s}) ,
\eeq
where $n_{\rm HII}$ is the total density of ions and electrons in the \ion{H}{2} region, $n_{\rm HI}$ is the atomic density in the neutral gas, $T_{\rm HII}$ and $T_{\rm HI}$ are the respective temperatures, and $f_{\rm turb}\geq 1$ is a factor quantifying the degree of pressure enhancement due to turbulence in the neutral gas.  Pressure equilibrium will be violated in non-steady-state, episodic accretion (\S~\ref{sec:timedep}).

Depending on the SED of the central source, the photoionization rate will be dominated by primary photoionizations or by secondary photoionizations carried out by photoelectrons \citep[e.g.,][]{Shull:85,Xu:91,Dalgarno:99}.  Taking into account only the ionization of hydrogen from the ground state, the total rate of photoionization in the annulus can be written $\dot N_{\rm ion}=f_{\rm ion} L /E_{\rm H}$, where $E_{\rm H}=13.6\textrm{ eV}$, $L=10^{40}L_{40}\textrm{ erg s}^{-1}$ is the luminosity shortward of $E_{\rm H}$, and $f_{\rm ion}\sim \langle E_{\rm H}/h\nu \rangle$ is the average fraction of the energy of an absorbed photon that goes into photoionization. For an almost fully ionized gas and a power law spectrum $f_\nu\propto \nu^{-\alpha}$, this fraction equals $f_{\rm ion}=(\alpha-1)/\alpha$ for $\alpha > 1$; in what follows, it should be borne in mind that $f_{\rm ion}$ depends on the shape of the SED. 

In a photoionization equilibrium $\dot N_{\rm ion}$ equals the total hydrogen recombination rate in the ionized gas $\dot N_{\rm rec} \sim \frac{4}{3}\pi r_{\rm ion}^3 \alpha_B(T_{\rm HII}) n_{\rm H^+} n_e $, where $\alpha_B(T)\sim 2.6\times 10^{-13}\ T_4^{-1} \textrm{ cm}^3\textrm{ s}^{-1}$ is the approximate hydrogen recombination coefficient in the on-the-spot approximation at temperatures $10^4\textrm{ K}\lesssim T<10^5\textrm{ K}$ \citep[][quoted in \citealt{Jappsen:08}]{Ferland:92}, and $n_{\rm H^+}\sim n_e\sim \frac{1}{2} n_{\rm HII}$ are the ion and electron densities, respectively.  Though not entirely justified in Str\"omgren spheres, the on-the-spot approximation is reasonable unless the density profile within the ionized region is sharply peaked toward the center \citep{Ritzerveld:05}, as is assumed not to be the case here.

\begin{figure}
\begin{center}
\includegraphics[width=3.0in]{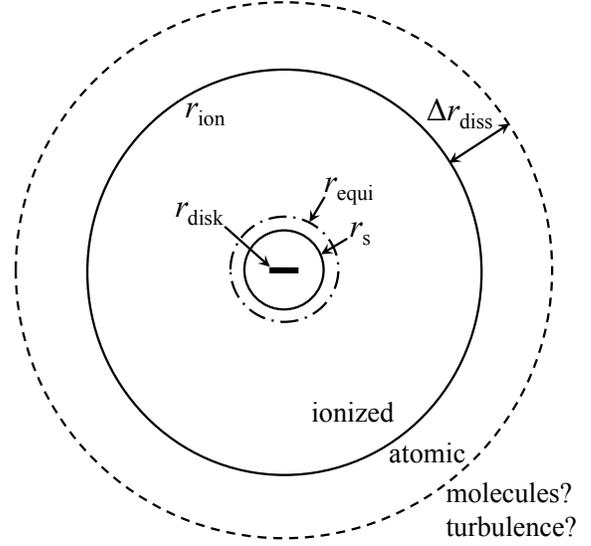}
\end{center}
\caption{Schematic representation (not to scale) of the structure of the \ion{H}{2} region surrounding an accreting seed black hole. For normal protogalactic densities, the radius of the \ion{H}{2} region $r_{\rm ion}$  (eq.~[\ref{eq:rion_ell}]) is a few orders of magnitude larger than the sonic radius $r_{\rm s}$ (eq.~[\ref{eq:rs}]). Gas is in local thermal and statistical equilibrium outside the radius $r_{\rm equi}$ (\S~\ref{sec:rs_conditions}), which may be slightly larger or smaller than $r_{\rm s}$.  The ionized gas contains a small, nonzero neutral fraction (eq.~[\ref{eq:chi_hydrogen}]), which gives rise to photoionization radiation pressure. The ionized region is surrounded by a shell of atomic gas of thickness $\Delta r_{\rm diss}$ where molecule photodissociation is efficient (\S~\ref{sec:photodissociation}). The outer, molecular shell may be supersonically turbulent.}
\label{fig:drawing}
\end{figure}

Equating the ionization rate to the recombination rate we obtain 
\beq
\label{eq:rion}
r_{\rm ion} \sim 4.4\times 10^{18}\ \frac{f_{\rm ion}^{1/3} ~ L_{40}^{1/3}~T_{{\rm HII},4.7} }{f_{\rm turb}^{2/3} ~ n_5^{2/3} ~ T_{{\rm HI},3.7}^{2/3}}\ \textrm{ cm} ,
\eeq
where $T_{\rm HII}=5\times 10^4\ T_{{\rm HII},4.7}\textrm{ K}$, $T_{\rm HI}=5\times10^3\ T_{{\rm HI},3.7}\textrm{ K}$, and $n_{\rm HI}=10^5\ n_5 \textrm{ cm}^{-3}$ (we justify our choice of the reference density in \S~\ref{sec:density} below).  If we express the luminosity in terms of the dimensionless ratio $\ell\equiv L/(4\pi GM_{\rm BH} m_p c/\sigma_{\rm T})$ of the luminosity to the Eddington luminosity (here and henceforth, for Thomson scattering), the radius of the \ion{H}{2} region becomes
\beq
\label{eq:rion_ell}
r_{\rm ion} \sim 4.7\times 10^{18}\ \frac{\ell^{1/3}~ f_{\rm ion}^{1/3} ~M_2^{1/3}~ T_{{\rm HII},4.7}}{ f_{\rm turb}^{2/3}~n_5^{2/3} ~ T_{{\rm HI},3.7}^{2/3}}\ \textrm{ cm} . 
\eeq
The ionization radius is normally much larger than the sonic radius,
\beq
\label{eq:rion_over_rs}
\frac{r_{\rm ion}}{r_{\rm s}} \sim 2\times10^4\ \frac{\ell^{1/3}~ f_{\rm ion}^{1/3}~ T_{{\rm HII},4.7}~T_{{\rm s},5}}{ f_{\rm turb}^{2/3}~n_5^{2/3} ~ T_{{\rm HI},3.7}^{2/3}~M_2^{2/3}} .
\eeq
In Figure \ref{fig:drawing} we provide a schematic illustration of the structure of the \ion{H}{2} region surrounding an accreting seed black hole.

Self-gravity of the gas inside the \ion{H}{2} region is negligible compared to that of the black hole, $\frac{4}{3}\pi r^3 n_{\rm H^+} m_p \ll M_{\rm BH}$, when $\ell \ll 0.054 \   f_{\rm ion}^{-1} f_{\rm turb} n_5 T_{{\rm HI},3.7} T_{{\rm HII},4.7}^{-2}  (r/r_{\rm ion})^{-3}$ which is satisfied for $r \ll r_{\rm ion}$, but need not always be true at the very edge of the \ion{H}{2} region.  There, however, the combined gravity of the black hole and the gas contained within $r_{\rm ion}$ are negligible compared to gas pressure gradients; the ratio of the Jeans length inside the \ion{H}{2} region to its radius is normally much larger than unity, $\lambda_{\rm J}/r_{\rm ion}\sim 28~  f_{\rm ion}^{-1/3} f_{\rm turb}^{1/6} \ell^{-1/3} n_5^{1/6} T_{{\rm HI},3.7}^{1/6} M_2^{-1/3}$, and thus self-gravity of the ionized gas can be ignored.

\subsection{Conditions at the Sonic Radius}
\label{sec:rs_conditions}

We will assume throughout that the gas is almost fully ionized at the sonic radius, i.e., that $r_{\rm s}\ll r_{\rm ion}$. This condition places an upper limit on the density of the neutral gas just outside the \ion{H}{2} region, $n \ll n_{\rm max,ion}$, where
\beq
\label{eq:n_optically_thin}
n_{\rm max,ion} = 2.7\times 10^{11} \frac{f_{\rm ion}^{1/2} ~ \ell^{1/2} ~T_{{\rm HII},4.7}^{3/2}~ T_{{\rm s},5}^{3/2}}{ f_{\rm turb} ~ T_{{\rm HI},3.7} ~ M_2} \textrm{ cm}^{-3} ,
\eeq
and we have assumed that $T_{{\rm HII},4.7}$ also represents the temperature of the ionized gas at the sonic radius. This shows that for an isotropic central radiation source, the gas at the sonic radius is guaranteed to be ionized unless the accreting gas has densities far in excess of those expected for the diffuse protogalactic medium and instead characteristic of self-gravitating, star-forming cores.

The central luminosity can be related to the mass accretion rate by substituting equation (\ref{eq:velocity_asymptotic}) into $\ell\equiv\epsilon r^2 v \mu n c \sigma_{\rm T}/GM_{\rm BH}$ to obtain the dimensionless luminosity for steady state accretion through the \ion{H}{2} region
\beq
\label{eq:ell_bondi}
\ell_{\rm s.s.} \sim 0.001~ \frac{\epsilon_{-1} ~f_{\rm turb}~\Upsilon ~M_2 ~n_5 ~T_{{\rm HI},3.7}}{
T_{{\rm HII},4.7}~ T_{\rm s,5}^{3/2}} ,
\eeq
where $\epsilon=0.1\epsilon_{-1}$ is the radiative efficiency.  Evidently, the mass accretion can be strongly suppressed by the heating of the ionized gas near the sonic radius. The accretion rate is much smaller than the Bondi accretion rate $\dot M_{\rm Bondi}=e^{3/2} \pi G^2M_{\rm BH}^2m_p n_{\rm HI}/c_{\rm s,HI}^3$ that would be calculated ignoring photoionization and photoheating altogether, i.e., for radiatively-inefficient accretion
\beq
\label{eq:bondi_naive}
\ell_{\rm Bondi} \sim 5.2\ \frac{ \epsilon_{-1}~ M_{\rm 2}~ n_5}{ f_{\rm turb}^{3/2}~T_{{\rm HI},3.7}^{3/2}}
\ \ \ \ 
(\textrm{no photoheating/ionization}) ,
\eeq
where  have assumed an isothermal equation of state.

To justify our assumption in \S~\ref{sec:photoionized_accretion} that the photoionized gas is in local thermal and statistical equilibrium, we calculate the ratio of the Bremsstrahlung cooling time $t_{\rm Brems}\sim 2.5\times10^{11} T^{1/2} n_{\rm H}^{-1}\textrm{ s}$ at the sonic radius to the inflow time at the sonic radius 
\beq
t_{\rm s}\sim \frac{r_{\rm s}}{c_{\rm s}(r_{\rm s})}\sim 1.45~ \frac{M_2}{T_{{\rm s},5}^{3/2}}\textrm{ yr} ,
\eeq 
to obtain
\beq
\label{eq:time_ratio_brems}
\frac{t_{\rm Brems}}{t_{\rm s}}\sim 77~\frac{ T_{{\rm s},5}^2 ~T_{{\rm HII},4.7}}{f_{\rm turb} ~\Upsilon~T_{{\rm HI},3.7} ~n_5~ M_2} .
\eeq
Cooling due to the recombination of \ion{H}{2} and \ion{He}{3} is comparable to and only slightly stronger than that due to Bremsstrahlung at $T_{\rm s}\lesssim 2\times10^5\textrm{ K}$.

We also estimate the ratio of the Compton heating time at the sonic radius  $t_{\rm C}\sim 0.0675\ G M \mu^2 m_e m_p c / \ell k^2T_{\rm s} T_{\rm C} $, where $T_{\rm C}\sim 10^7 T_{{\rm C},7}\textrm{ K}$ is the Compton temperature, to the inflow time \citep[see also Fig.~2 in][]{Sazonov:05}, 
\beq
\label{eq:time_ratio_compton}
\frac{t_{\rm C}}{t_{\rm s}}\sim 0.01~\frac{T_{{\rm s},5}^{1/2}}{\ell~ T_{{\rm C},7}} .
\eeq
At the relatively high gas densities considered here, which are required for a rapid growth of seed massive black holes in protogalaxies, the Compton cooling of the photoionized gas by the cosmic microwave background photons \citep[see, e.g.,][]{Ricotti:08} can be ignored. 

Substituting equation (\ref{eq:ell_bondi}) in equation (\ref{eq:time_ratio_compton}) we obtain for ratio of the Compton heating time to the infall time
\beq
\label{eq:time_ratio_compton_without_ell}
\frac{t_{\rm C}}{t_{\rm s}}\sim 11~\frac{ T_{{\rm s},5}^2 ~T_{{\rm HII},4.7}}{\epsilon_{-1}~f_{\rm turb}~\Upsilon~ n_5~ M_2 ~T_{{\rm C},7}~ T_{{\rm HII},3.7}} .
\eeq
Equations (\ref{eq:time_ratio_brems}) and (\ref{eq:time_ratio_compton_without_ell}) suggest that heating and cooling times at the sonic radius can be longer than the inflow time and that 
the ionized gas may not be in local thermal and statistical equilibrium at all radii $r\gg r_{\rm s}$, especially if the accretion occurs below the Eddington-limited rate. However, the ionized gas should be in equilibrium at only a somewhat larger radius because the infall time increases rapidly with radius, $r/v \propto r^3$. 

So far we have ignored the dependence of the temperature at the sonic radius on the radiative efficiency, which as Figure \ref{fig:tofxi} shows, can be strong.  The precise form of the function $T_{\rm s}(\epsilon)$, defined by the intersection of the $T_{\rm eq}(\Xi)$ curve and the $T_{\rm s}(\Xi)$ line, depends sensitively on the metallicity of the gas and on the SED of the central source and we do not attempt to model it in general.  If the dependence can be approximated with a power low in a range of efficiencies, $T_{\rm s}\propto \epsilon^\delta$, from equations (\ref{eq:time_ratio_brems}), (\ref{eq:ell_bondi}), and (\ref{eq:time_ratio_compton_without_ell}) we obtain 
\beq
\frac{t_{\rm Brems}}{t_{\rm s}}\propto \epsilon^{2\delta}, 
\ \ \ \ell \propto \epsilon^{1-3\delta/2}, \ \ \  \frac{t_{\rm C}}{t_{\rm s}}\propto \epsilon^{2\delta-1} .
\eeq
If, e.g., $\delta\sim 1$ for $\epsilon\gtrsim 0.1$, we find that with an increasing efficiency it becomes more difficult for the gas to achieve local thermal equilibrium at the sonic radius, but if the equilibrium is achieved, the central luminosity, and especially the accretion rate that is proportional to $\dot M\propto \ell/\epsilon\propto \epsilon^{-3\delta/2}$, decrease with increasing efficiency.

\citet{Ciotti:97,Ciotti:01,Ciotti:07} and \citet{Sazonov:05} have studied spherically-symmetric accretion of hot interstellar medium onto an X-ray quasar in an elliptical galaxy and have identified a limit cycle driven by Compton heating. The quasar heats the interstellar medium to temperatures exceeding the virial temperature, which leads to an outflow and quenching of central accretion. This model differs from ours in that a metal-enriched environment is assumed, so that photoionization equilibrium temperature reaches the Compton temperature already at $\Xi\sim50$, whereas in our metal-poor model the transition to the Compton temperature occurs at higher values of the photoionization parameter, $\Xi\sim 10^3$.  Therefore, in the model of \citet{Sazonov:05}, the ionized gas can reach the Compton temperature well outside the sonic radius.  Another crucial difference is the assumed ionized gas density profile far from the sonic radius: the hot gas surrounding a quasar was assumed to be hydrostatically confined by the gravity of the host galactic stellar spheroid such that its density declines steeply with radius, $n_{\rm HII}\propto r^{-2}$, which implies that $\Xi$ is roughly independent of radius and so the Compton-heated equilibrium can exist at arbitrarily large radii.  In our model, since the \ion{H}{2} region surrounding a seed black hole is confined by external pressure  at radii $r\gg r_{\rm s}$ and is roughly isothermal (consistent with the very weak dependence of $T_{\rm eq}$ on the dimensionless ionization parameter for $10\lesssim \Xi \lesssim 10^3$; see Fig.~\ref{fig:tofxi}), the density inside it is approximately independent of radius. Then $\Xi\propto r^{-2}$, and the gas is progressively farther from being able to heat the Compton temperature at radii much larger than the sonic radius.

\subsection{Protogalactic Density at the Edge of the \ion{H}{2} Region}
\label{sec:density}

We will now pause our investigation of radiative feedback effects to clarify our choice of the density of the ambient neutral gas surrounding the \ion{H}{2} region.  Rapidly growing protogalaxies contain gas with a wide range of densities, and thus care must be taken to appropriately specify the gas density on scales relevant for regulation of accretion onto a seed black hole by radiative feedback effects. In the simplest picture, central gravitational collapse of the gas at the center of a protogalaxy produces a distribution in which gas density is spherically symmetric and a function of radius only, $n(r)$. If the gas is approximately isothermal and quasi-hydrostatic in the presence of turbulence, $n(r)\propto r^{-2}$.  If we ignore the gravity of any seed black hole and let $f_{\rm turb}$ denote a radius-independent turbulent pressure enhancement, the density scales as
\bea
\label{eq:density_cusp}
n(r) &\sim& \frac{f_{\rm turb} ~c_{\rm s}^2}{2\pi ~G ~m_p ~r^2}\nonumber\\
&\sim& 6\times10^4~ f_{\rm turb} ~T_{\rm HI,3.7} \left(\frac{r}{1\textrm{ pc}}\right)^{-2}\textrm{ cm}^{-3} .
\eea
Equation (\ref{eq:density_cusp}) is a good approximation to the density ``cusp'' profile resulting from central gravitational collapse in a $M_{\rm halo}\sim 10^8M_\odot$ cosmological halo in the simulations of \citet{Bromm:03} and in the simulations of \citet{Wise:08}, who found turbulent Mach numbers ${\cal M}\sim 3$ in the cusp, which would imply $f_{\rm turb}\sim 10$.  

If a seed black hole is located near the center of the density cusp, radiative effects may prevent the gravitational collapse from proceeding to the arbitrarily large densities. We will here assume that radiative effects prevent collapse at radii smaller than the radius of the \ion{H}{2} region, i.e., that the density profile of the neutral gas has a ``core'' on scales $\sim r_{\rm ion}$ such that density within the \ion{H}{2} is roughly uniform far from the sonic radius.  For self-consistency, we substitute $n=n(r_{\rm ion})$ from equation (\ref{eq:density_cusp}) in equation (\ref{eq:rion_ell}) and solve for $r_{\rm ion}$, and in turn for $n(r_{\rm ion})$, to obtain
\bea
\label{eq:r_dens_self_consistent}
r_{\rm ion,cusp} &\sim& 3.8\times 10^{17}~\frac{ f_{\rm turb}^4  ~T_{{\rm HI},3.7}^4}{f_{\rm ion} ~\ell~ T_{{\rm HII},4.7}^{3}~ M_2} \textrm{ cm} ,\nonumber\\
n_{\rm cusp} &\sim& 4.5\times10^6~ \frac{f_{\rm ion}^2 ~\ell^2 ~ T_{{\rm HII},4.7}^6~ M_2^2 }{ f_{\rm turb}^7~ T_{{\rm HI},3.7}^7} \textrm{ cm}^{-3}.
\eea
These estimates are but crude self-consistency conditions and suffer from a strong sensitivity to the turbulent Mach number and other parameters.  The estimate of gas density at the edge of the \ion{H}{2} region given in equation (\ref{eq:r_dens_self_consistent}) can only be used as a rough guide for the range of densities that should be addressed in the ensuing analysis.   Our choice of the reference density, $10^5\textrm{ cm}^{-3}$, is compatible with the self-consistency conditions in a protogalaxy in a cosmological halo of $M_{\rm halo}\sim 10^8M_\odot$ for $f_{\rm ion}\sim \frac{1}{3}$, ignoring any recent supernova activity in the center of the halo, which can drastically reduce the central density \citep[see, e.g.,][]{Wada:03,Kitayama:05,Wise:08b}.

Equations (\ref{eq:r_dens_self_consistent}) also suggest that ``minihalos'' with masses $M_{\rm halo}\sim 10^6M_\odot$ and Mach numbers ${\cal M}\gtrsim 1$ will be fully ionized out to $r\gtrsim 100\textrm{ pc}$ if, somehow, $\ell_{\rm max}\sim1$ is realized at an early instant prior to the expansion of the \ion{H}{2} region, and the interior of the \ion{H}{2} region has not had chance to heat beyond $10^4\textrm{ K}$.  This is seen in the simulations of radiative feedback during accretion onto seed black holes in cosmological minihalos of \citet{Alvarez:08}.

\subsection{Continuum Radiation Pressure}
\label{sec:prad_continuum}

Here we estimate the effects of the continuum radiation pressure in the interior of the \ion{H}{2} region. The continuum radiation pressure acceleration is $a_{\rm rad} = \bar \sigma_{\rm H} L \chi_{\rm HI} /4\pi r^2  m_p c$,
where $\chi_{\rm HI}$ is the abundance of neutral hydrogen in the ionized gas, and $\bar\sigma_{\rm H}=F^{-1} \int \sigma_{{\rm H},\nu} F_\nu d\nu$ is the frequency-averaged mean absorption cross section. In a highly ionized gas, most of the photoelectron energy goes into heating, so in ionization balance
\beq
 \chi_{\rm HI} =
\frac{\alpha_{B}(T_{\rm HII})}{L ~\tilde \sigma_{\rm H}/(4\pi~ r^2 ~n_e~ E_{\rm H}) + \alpha_{\rm ion}(T_{\rm HII}) } ,
\eeq
where $\tilde \sigma_{\rm H}=F^{-1} E_{\rm H}\int (h\nu)^{-1} \sigma_{{\rm H},\nu} F_\nu d\nu$ and $\alpha_{\rm ion}(T)$ is the collisional ionization rate. At temperatures $\ll 10^5\textrm{ K}$, photoionization typically dominates collisional ionization.  The neutral hydrogen abundance then becomes
\beq
\label{eq:chi_hydrogen}
 \chi_{\rm HI} \sim 7.6\times 10^{-4}\ \frac{f_{\rm ion}^{2/3}}{\ell^{1/3}~f_{\rm turb}^{1/3}~ M_2^{1/3} ~n_5^{1/3} ~T_{{\rm HI},3.7}^{1/3}} \left(\frac{r}{r_{\rm ion}}\right)^2 ,
\eeq
which implies that at radii not much smaller than $r_{\rm ion}$, the neutral abundance is $\gtrsim 10^{-6}$, and in this regime, 
since $\bar{\sigma}_{\rm H} \sim 10^{6} \sigma_{\rm T}$,
radiation pressure due to photoionization exceeds that due to Thomson scattering.  The acceleration due to the former is then given by
\beq
a_{\rm rad} = \frac{\bar \sigma_{\rm H}}{\tilde\sigma_{\rm H}} ~ \frac{\alpha_{B}(T_{\rm HII})~  n_e ~E_{\rm H} }{m_p~ c} .
\eeq
Note that, unlike in the case of Thomson scattering, this is {\it independent} of distance from the black hole because here the abundance of absorbers increases with the square of the radius, $\chi_{\rm HI} \propto r^2$.  
The cross section ratio is $\bar \sigma_{\rm H}/\tilde \sigma_{\rm H}\sim (3+\alpha)/(2+\alpha)$ for power-law spectra $F_\nu\propto \nu^{-\alpha}$.  We compare $a_{\rm rad}$ to  the black hole's gravity $a_{\rm grav}= - GM_{\rm BH} / r^2$ to find,
\bea
\label{eq:cont_press_HII}
\frac{a_{\rm rad}}{|a_{\rm grav}|} &\sim& 980~ \frac{f_{\rm ion}^{2/3}~\ell^{2/3}}{f_{\rm turb}^{1/3} ~T_{{\rm HI},3.7}^{1/3} ~n_5^{1/3} ~M_2^{1/3}}  \left(\frac{r}{r_{\rm ion}}\right)^2
\eea
where we have assumed a fiducial $\alpha=1.5$ spectrum, implying $f_{\rm ion}\sim \frac{1}{3}$ and $\bar \sigma_{\rm H}/\tilde \sigma_{\rm H}\sim \frac{9}{7}$.
The relative importance of photoionization radiation pressure depends on distance, and dominates at larger radii. 
This shows that for a wide range of parameters, for near-Eddington accretion, the radiation pressure in the \ion{H}{2} region at radii $r\lesssim r_{\rm ion}$ greatly exceeds the black hole's gravity, which implies that steady state accretion may occur at a reduced rate. Inward accretion is still possible if a positive pressure gradient is set up to counteract the radiation pressure force.

To estimate the reduction of the steady-state accretion rate as a result of the photoionization radiation pressure in the \ion{H}{2} region, we simplify the mathematical problem by ignoring any heating of the gas near the sonic radius and setting $T_{\rm s}=T_{\rm HII}={\rm const}$; this simplification, as we shall see, will turn out to be unreasonably optimistic.  Then, after a change of variables,
\beq
w \equiv \frac{v}{c_{\rm s,HII}} , \ \ \ \ \ 
y \equiv  r \left(\frac{GM_{\rm BH}}{2c_{\rm s,HII}^2}\right)^{-1} \approx \frac{r}{r_{\rm s}} ,
\eeq
equation (\ref{eq:inflow}) describing momentum conservation in the accretion flow becomes 
\beq
\label{eq:flow_nondim}
\frac{dw}{dy}~(w^2-1) = w~ \left(\frac{2}{y} - \frac{2}{y^2} + \phi\right) ,
\eeq
where $\phi$ is a dimensionless parameter proportional to the photoionization radiation pressure acceleration,
\bea
\phi&\equiv&  \frac{GM_{\rm BH}}{2 c_{\rm s,HII}^4}~a_{\rm rad}\nonumber\\
&\sim& 2\times10^{-5} ~\frac{f_{\rm turb} ~ n_5 ~M_2 ~T_{{\rm HII},3.7}}{T_{{\rm HII},4.7}^4} .
\eea
Equation (\ref{eq:flow_nondim}), subject to the regularity condition imposed at the sonic radius, $w(y_{\rm s})=1$, where $y_{\rm s}= \phi^{-1} [(1+2\phi)^{1/2}-1]\approx 1$, can be solved in closed form in terms of the transcendental Lambert $W(x)$ function, defined implicitly via $x=We^W$, to obtain
\bea
w(y)&=&\sqrt{-W\left\{-\left(\frac{y}{y_{\rm s}}\right)^{-4}\exp\left[-\frac{2(\phi y^2+2)}{y}+\frac{2(\phi y_{\rm s}^2+2)}{y_{\rm s}}-1\right]\right\}} \nonumber\\
&\approx& y^{-2} \exp\left(-\frac{\phi y^2 - 3y/2 + 2}{y}\right) ,
\eea
where in the second line we applied the expansion $W(x)=x+{\cal O}(x^2)$.

For $\phi\ll 1$, the ratio of density $n\propto (r^2 v)^{-1} \propto (y^2 w)^{-1}$ at radius $r\gg r_{\rm s}$ to that at the sonic radius has asymptotic form
\beq
\label{eq:density_gradient}
\frac{n(r)}{n(r_{\rm s})} \sim e^{\phi ~r / r_{\rm s}} ~\left[\frac{n(r)}{n(r_{\rm s})}\right]_{\phi=0} . 
\eeq
This result is exact for isothermal accretion, but here, we adopt it as an optimistic estimate of the density reduction due to photoionization radiation pressure in the case of realistic, nonisothermal accretion.

We expect significant reduction in the accretion rate when $n_{\rm HII}(r_{\rm s}) \ll n_{\rm HII}(r_{\rm ion})$ or, according to equation (\ref{eq:density_gradient}), when the quantity
\beq
\label{eq:rad_force_parameter}
\phi~\frac{r_{\rm ion}}{r_{\rm s}} ~\sim~ 0.4~\frac{f_{\rm ion}^{1/3}~f_{\rm turb}^{1/3}~\ell^{1/3}~T_{{\rm HI},3.7}^{1/3}~T_{{\rm s},5}~n_5^{1/3}~M_2^{1/3}}{T_{{\rm HII},4.7}^3}
\eeq
is comparable to or greater than unity.
The dimensionless accretion rate can be eliminated by substituting equation (\ref{eq:ell_bondi}) that relates the accretion rate to the conditions at the sonic radius, to obtain
\beq
\phi~\frac{r_{\rm ion}}{r_{\rm s}} ~\sim~ 0.04~\frac{f_{\rm ion}^{1/3}~f_{\rm turb}^{1/3}~\Upsilon^{1/3}~\epsilon_{-1}^{1/3}~T_{{\rm HI},3.7}^{2/3}~T_{{\rm s},5}^{1/2}~n_5^{2/3}~M_2^{2/3}}{T_{{\rm HII},4.7}^{10/3}} .
\eeq
Thus we find that accretion will proceed in one of the two distinct regimes. When 
\beq
\phi \frac{r_{\rm ion}}{r_{\rm s}}\ll 1 ,
\eeq 
radiation pressure does not affect the accretion rate even if the combined gravitational and radiative acceleration is outward in parts of the \ion{H}{2} region.  However, when 
\beq
\phi \frac{r_{\rm ion}}{r_{\rm s}}\gtrsim 1 ,
\eeq 
the ionized gas assumes a density gradient at radii $r\gg r_{\rm s}$ such that the density increases outward, toward the edge of the \ion{H}{2} region.  In this case, central accretion is suppressed at least by the factor \beq
\ell \propto \exp\left(-\phi\frac{r_{\rm ion}}{r_{\rm s}}\right) , 
\eeq
and probably more, because the central drop in density will lead to a photoionization equilibrium at a higher temperature, thereby increasing gas temperature near the sonic radius.  Therefore, for $\phi r_{\rm ion}/r_{\rm s}\gtrsim 1$, radiation pressure acting on the ionized gas imposes a more restrictive limit on the accretion rate than the mere photoheating near the sonic radius. The two limits are complementary: while the photoheating limit is more restrictive for small black hole masses, ambient densities, and turbulent Mach numbers, the radiation pressure limit is more restrictive in the very opposite regime.  A combination of these two limits suggests that a steady state solution for near-Eddington accretion does not seem to exist.\footnote{\citet{Wang:06} concluded that radiation pressure was negligible because they erroneously assumed complete ionization ($\chi_{\rm HI}=0$).}

A fraction $f_{\rm res}<1$ of the luminosity emitted by the central source passes unabsorbed through the \ion{H}{2} region and reaches its edge, where most of it is absorbed in the partially ionized gas.  Momentum deposition by the residual luminosity leads to a density drop by the factor of $\exp(-\psi)$ where
\bea
\psi&\equiv& \frac{f_{\rm res}~L}{4\pi ~r_{\rm ion}~\gamma ~n_{\rm HII} ~k ~T_{\rm HII} ~c}\nonumber\\
&\sim&\left(0.033~\frac{f_{\rm res}~T_{{\rm HII},4.7}}{f_{\rm ion}~ T_{{\rm s},5}} \right)~\phi\frac{r_{\rm ion}}{r_{\rm s}} .
\eea
Since $T_{\rm s}\geq T_{\rm HII}$, radiation pressure in the partially-ionized edge of the \ion{H}{2} region is always less important than in its interior.  

\subsection{Accretion at Low Radiative Efficiencies}
\label{sec:radiatively_inefficient}

In this work we focus entirely on the regime in which central accretion is radiatively efficient, e.g., $\epsilon\gg 0.01$, as it should be the case when the radiation is emitted by a geometrically thin disk.  In this regime, heating of the gas at the sonic radius to $\gtrsim 10^5\textrm{ K}$, and perhaps even to the Compton temperature at high metallicities, seems inevitable.  We hope to generalize our analysis to radiatively inefficient accretion, which should be accompanied by outflows and shadowing, in a subsequent investigation. Here, we will make only a brief mention of this alternate regime. 

If we ignore the mechanical and radiative effects of the outflows and the shadowing, which is an implausible and inconsistent assumption, a change in the thermodynamic behavior occurs at $\epsilon\lesssim 10^{-3}$, namely, at such low efficiencies the gas at the sonic radius and throughout the \ion{H}{2} region can remain at temperatures well below $10^5\textrm{ K}$ \citep[see, e.g.,][]{Buff:74,Hatchett:76,Krolik:81,Kallman:82,Krolik:84,Lepp:85,Donahue:91}.  If, e.g., $T_{\rm s}\sim T_{\rm HII} \sim 10^4\textrm{ K}$ and $\Upsilon\sim 1$, we find that, still, under a wide range of conditions characteristic of protogalactic clouds, an extended ($r_{\rm ion}\gg r_{\rm s}$) \ion{H}{2} region exists around the seed black hole.  In this case, photoionization radiation pressure suppresses central accretion when $\phi r_{\rm ion}/r_{\rm s}\gtrsim 1$, which can be expressed as a condition on the radiative efficiency; the suppression takes place when $\epsilon\gtrsim\epsilon_{\rm crit}$ where
\beq
\epsilon_{\rm crit} \sim \frac{0.005}{ f_{\rm ion}~ f_{\rm turb}^2 ~T_{{\rm HI},3.7}^2~ n_5^2~M_2^2} .
\eeq
For $\epsilon\gtrsim \epsilon_{\rm crit}$, the ionized gas may not be able to heat above $10^4\textrm{ K}$, but the photoionization radiation pressure suppresses accretion well below the rate $\ell_{\rm s.s.}$ derived in equation (\ref{eq:ell_bondi}).
For $\epsilon\ll \epsilon_{\rm crit}$, accretion is limited by conditions at the sonic radius (\S~\ref{sec:rs_conditions}). The accretion rate may become close to the ``Bondi'' rate calculated ignoring radiative effects altogether (eq.~[\ref{eq:bondi_naive}]) if the neutral gas surrounding the \ion{H}{2} region is weakly supersonically turbulent, but the ionized gas inside it is not.  

We would like to reiterate that the analysis presented in this subsection is somewhat unrealistic, because radiatively-inefficient accretion flows are probably accompanied by outflows that can radically alter the hydrodynamic, thermodynamic, and chemical structure of the region surrounding a seed black hole.  Also, the radiation field produced by a radiatively-inefficient accretion flow is potentially highly anisotropic.

\subsection{${\rm Ly}\alpha$ Radiation Pressure}
\label{sec:prad_lyalpha}

Another source of pressure is from the Ly$\alpha$ photons that are produced throughout the \ion{H}{2} region and can become trapped within the ionized region and the surrounding neutral shell \citep[see, e.g.,][and references therein]{McKee:07}.  Let $f_{{\rm Ly}\alpha}\sim \frac{2}{3}$ \citep[e.g.,][]{Osterbrock:89}  denote the fraction of the total power emitted by the central source above $E_{\rm H}$ that is reprocessed into Ly$\alpha$ photons.  Consider a shell of neutral gas at radii $r_{\rm ion}< r < r_{\rm ion}+\Delta r_{\rm shell}$, where $\Delta r_{\rm shell}$ denotes the thickness of the shell. For shells with $\Delta r_{\rm shell} \lesssim r_{\rm ion}$, the line-center optical depth of the shell $\tau_0 = \sigma_0(T) n_{\rm H} \Delta r_{\rm shell}$, where $\sigma_0(T)=5.9\times10^{-14} T_4^{-1/2}\textrm{ cm}^2$, is  large
\beq
\tau_0 \sim 4\times 10^{10} \ \frac{f_{\rm ion}^{1/3}~\ell^{1/3}~ M_2^{1/3}~ n_{\rm 5}^{1/3}~ T_{{\rm HII},4.7}}{f_{\rm turb}^{2/3}~ T_{{\rm HI},3.7}^{7/6}} \left(\frac{\Delta r_{\rm shell}}{r_{\rm ion}}\right) .
\eeq
Since the shell is very optically thick, a photon injected at the edge ($r=r_{\rm ion}$) near line center will scatter across the ionized region many times before escaping the shell. The ionized gas within the \ion{H}{2} region has some residual optical depth due to the presence of a small neutral fraction. Under the conditions considered here, the ionized gas is optically thick to the photons in the core of the Ly$\alpha$ line, and is marginally optically thin to the photons in the wings.  We ignore this complication and treat the ionized gas as optically thin.   

To estimate the number of times a photon injected at the edge of the neutral gas crosses the \ion{H}{2} region, we assume that $\Delta r_{\rm shell}\lesssim r_{\rm ion}$, and carry out a simple Monte Carlo calculation of a photon's frequency diffusion before it escapes the shell.   For this, we employ the accurate expressions for the transmission coefficient and the reflection frequency redistribution function at the shell edge that \citet{Hansen:06} obtained from Monte-Carlo resonant line scattering calculations.  In the dust-free limit, the number of times a photon injected near the line core reflects against the walls of the ionized region is accurately approximated with
\beq
N_{\rm reflect} \approx 0.609~ (a~\tau_0)^{0.659} ~ e^{-0.00607~ [\ln(2120.0 ~a~\tau_0 )]^2} ,
\eeq
where  $a(T)=\nu_{\rm L}/2\nu_{\rm Dop}=4.72\times10^{-4} T_4^{-1/2}$ is the ratio of the natural to the Doppler line width.  
This is close to the asymptotic formula due to \citet{Adams:72,Adams:75}, 
$N_{\rm reflect} \sim 15\ (\tau_0/10^{5.5})^{1/3}$, in the optically thick limit \citep[quoted from][]{Dijkstra:08}.
At $T=10^4\textrm{ K}$, for $\tau_0=(10^9,10^{10},10^{11})$, the numbers of reflections take values $N_{\rm reflect}\sim (250,610,1400)$.  Rayleigh scattering contributes to the opacity negligibly for $\tau_0\lesssim 10^{10}$.  Destruction of Ly$\alpha$ photons by stimulated two photon emission is negligible in the range of optical depths that we consider. Destruction by H$^-$ ions in the partially ionized shell surrounding the \ion{H}{2} region is potentially important (see \S~\ref{sec:photodissociation}), but because of significant uncertainties, we ignore it and adopt Adams' formula in what follows. 

If the radiation pressure drives a coherent expansion (outflow) in the neutral shell, this may promote photon escape and reduce $P_{{\rm Ly}\alpha}/P_{\rm gas}$.  If the outflow contains a velocity gradient of magnitude $\Delta v$, \citet[][see also \citealt{Bithell:90}]{Bonilha:79} estimate that $N_{\rm scatter}$ is
reduced by a factor $\sim (1+0.04\ |\Delta v/v_{\rm Dop}|^{3/2})^{-1}$, where $v_{\rm Dop}\sim c_{\rm s}\sim 10\textrm{ km s}^{-1}$ is the thermal or turbulent Doppler velocity. The same reduction factor should apply to $N_{\rm reflect}$. Thus, even for a highly supersonic outflow $\Delta v/v_{\rm Dop} \sim 8$, the number of reflections to escape and the Ly$\alpha$ radiation pressure to which it is proportional will drop by only a half.  

The energy density in Ly$\alpha$ photons can be approximated with \citep[see, e.g.,][]{Bithell:90,Haehnelt:95}
\beq
U_{{\rm Ly}\alpha} \sim  \frac{3f_{{\rm Ly}\alpha}~ L~ N_{\rm reflect}}{4\pi (r+\Delta r_{\rm shell})^2 c} .
\eeq
From this, the radiation pressure $P_{{\rm Ly}\alpha}=\frac{1}{3} U_{{\rm Ly}\alpha}$ scale height is $\sim \frac{1}{2} r_{\rm ion}$.  At depth equal to one scale height we have, with the fiducial choice $f_{{\rm Ly}\alpha}=\frac{2}{3}$,
\beq
\label{eq:p_ratio_lyalpha}
\frac{P_{{\rm Ly}\alpha}}{P_{\rm gas}} \sim 3.8\ \frac{\ell^{4/9}~ f_{\rm turb}^{10/9}~ M_2^{4/9}~ n_5^{4/9} }{f_{\rm ion}^{5/9} ~T_{{\rm HI},3.7}^{1/18} ~T_{{\rm HII},4.7}^{5/3}} .
\eeq
For near-Eddington accretion the Ly$\alpha$ pressure can significantly exceed the gas pressure and impart an outward impulse to the neutral gas surrounding the \ion{H}{2} region.  This result appears to imply that even at relatively small $\ell$, the pressure of the trapped resonance line radiation remains in excess of the gas pressure.  

If we speculatively require $P_{{\rm Ly}\alpha}<P_{\rm gas}$ for strictly stationary accretion, we invert equation (\ref{eq:p_ratio_lyalpha}) to derive an upper limit on the accretion rate, $\ell<\ell_{{\rm crit,Ly}\alpha}$, where
\beq
\ell_{{\rm crit,Ly}\alpha}= 0.05~\frac{f_{\rm ion}^{5/4} ~T_{{\rm HI},3.7}^{1/8}~ T_{{\rm HII},4.7}^{15/4}}{f_{\rm turb}^{5/2}~M_2 ~n_5} .
\eeq
Because of the strong dependence on $T_{\rm HII}$ and our not having sought, within the confines of the present work, a self-consistent dynamical solution for gas flow in the presence of resonance line scattering radiation, we are not able determine whether the continuum photoionization pressure or the resonance line scattering is more constraining to the maximum accretion rate for strictly stationary quasiradial accretion that can be achieved.  Additional sources of uncertainty affecting any attempt to estimate the impact of resonance line radiation pressure are the topology of the density field beyond the spherically symmetric approximation, as the radiation may ``leak out'' even through a relatively small hole in the neutral shell \citep[see, e.g.,][]{Hansen:06,McKee:07} and the dynamical stability of shells accelerated by resonance line radiation \citep{Mathews:92}.

\section{Episodic Accretion}
\label{sec:timedep}

\subsection{Mechanics of Inflow and Outflow}
\label{sec:inflow_outflow}

Here we  revoke the assumption of strictly stationary flow and attempt to describe a sequence of periodically recurring stages in the seed black hole's accretion cycle.  Let $\ell_{\rm max}<1$ denote the peak luminosity in units of the Eddington luminosity. The peak luminosity is reached due a sudden infall of material; the thermodynamic state of the infalling gas is assumed to be set by a central luminosity well below the peak luminosity.  The infalling gas reacts to the sudden rise in central luminosity on a finite time scale. 
Facing a sudden rise in the central luminosity, the accreting gas heats by photoionization and gets accelerated outward by the radiation pressure. The photoionization heating time $t_{\rm photo}=4\pi r^2(\gamma-1)^{-1} kT_{\rm HII}/\frac{1}{2}\chi_{\rm H}\bar\sigma_{\rm H} L$, where $\bar \sigma_{\rm H}$ is the average photoionization cross section and $\chi_{\rm H}$ is the neutral hydrogen abundance (see \S~\ref{sec:prad_continuum}), is given by
\beq
t_{\rm photo}\sim 2.9~ \frac{T_{{\rm HII},4.7}^3}{f_{\rm turb}~f_{{\rm epi},1}~n_5~T_{{\rm HI},3.7}}\textrm{ yr} ,
\eeq
where we have included a potentially large density enhancement $f_{\rm epi}=10~f_{\rm epi,1}\geq 1$, due to, e.g., infall or a lack of pressure equilibrium, over the steady-state density of the \ion{H}{2} region.
The photoionization heating time scale is thus rather short and comparable to the inflow time at $r\sim r_{\rm s}$.  

Following the sudden increase in central luminosity, material in the outer part of the \ion{H}{2} region, $r\sim r_{\rm ion}$, is accelerated outward by the radiation pressure if  $a_{\rm rad} \gtrsim P_{\rm gas,HII}/r \mu m_p n_{\rm HII}$, which implies the condition (cf.~eq.~[\ref{eq:rad_force_parameter}])
\beq
3.3~\frac{f_{\rm ion}^{1/3}~f_{\rm turb}^{1/3}~f_{{\rm epi},1}~\ell_{\rm max}^{1/3}~T_{{\rm HI},3.7}^{1/3}~n_5^{1/3}~M_2^{1/3}}{T_{{\rm HII},4.7}^2} \gtrsim 1 .
\eeq
Photoionization heating and radiation pressure will act to erase the infall-induced central density enhancement and drive the central density and the accretion rate down, toward the limits imposed by thermodynamics (\S~\ref{sec:rs_conditions}) and radiation pressure (\S~\ref{sec:prad_continuum}). 
The accretion flow remains directed inward close to the sonic radius, but far from the sonic radius, the inflow gives way to an outflow.  

If the outflowing material reaches radii $\sim r_{\rm ion}$, it encounters a shell of denser hot gas that has been photoevaporated from edge of the \ion{H}{2} region. The hot shell is overpressured by a factor $\sim T_{\rm HII}/f_{\rm turb}T_{\rm HI}$ and is expanding in both radial directions.  As the central density and luminosity drop, radiation pressure is not able anymore to accelerate the ionized gas against the positive pressure gradient in the exterior of the \ion{H}{2} region. Then the pressure gradient resulting from the rapid decline of radiation pressure and from the photoevaporative heating accelerates the gas near the edge of the \ion{H}{2} region inward, in an implosion back toward the black hole. The gravitational acceleration is  subdominant compared to the pressure gradient acceleration, $|a_{\rm pres}|\gtrsim a_{\rm rad,max} \gg |a_{\rm grav}|$ (see eq.~[\ref{eq:cont_press_HII}]) until the infalling gas approaches to within a few sonic radii from the black hole, where, by definition, gravity becomes competitive with pressure.   The returning gas acquires a velocity similar to the local sound speed of the photoevaporated gas, which is perhaps somewhat larger than $c_{\rm s,HII}(r_{\rm ion})$, and returns to the vicinity of the black hole on a time scale 
\bea
\label{eq:time_return}
t_{\rm return}&\sim& \frac{r_{\rm ion}}{c_{\rm s,HII}}
\nonumber\\
&\sim& 4\times 10^4 ~ \frac{\ell_{\rm max}^{1/3} ~f_{\rm ion}^{1/3}  ~M_2^{1/3} T_{{\rm HII},4.7}^{1/2}}{f_{\rm turb}^{2/3}~  n_5^{2/3}~ T_{{\rm HI},3.7}^{2/3} }\textrm{ yr} .
\eea
The return completes the accretion cycle.
The return time may be shorter by the factor of a few than this estimate if the overpressuring during to photoevaporation, which we ignored in equation (\ref{eq:time_return}), is taken into account.  

The pressure of the Ly$\alpha$ scattering radiation trapped within, and in the neighborhood, of the \ion{H}{2} region may further reduce the average accretion rate by imparting an outward impulse to partially ionized gas and thus possibly significantly extending the return time $t_{\rm return}$.  In view of the lingering concerns regarding the transfer of Ly$\alpha$ radiation that we have raised in \S~\ref{sec:prad_lyalpha}, we ignore it in the following estimate of the duty cycle.

\subsection{Estimates of the Duty Cycle}
\label{sec:duty_cycle}

If, ultimately, after the central density and luminosity have decreased substantially, thermodynamics imposes  the most stringent limit on the accretion rate, we might expect that the luminosity drop toward the steady-state rate $\sim \ell_{\rm s.s.}$ derived in equation (\ref{eq:ell_bondi}).  Because of the sensitivity to the various parameters of our model, we are not able to generally determine whether the luminosity reaches $\ell_{\rm s.s.}$ in time $t_{\rm return}$ separating consecutive infall episodes.  For the simplicity of the remaining analysis of episodic accretion, we assume that the minimum luminosity is indeed $\sim \ell_{\rm s.s.}$ and that the decline is exponential, 
\beq
\ell(t) = \ell_{\rm max} \left(\frac{\ell_{\rm s.s.}}{\ell_{\rm max}}\right)^{t/t_{\rm return}} .
\eeq
Then, the time-average luminosity is given by
\beq
\label{eq:ellbar_episodic}
\langle \ell \rangle = \frac{\ell_{\rm max}}{\ln(\ell_{\rm max}/\ell_{\rm s.s.})} ,
\eeq
and the duty cycle, which we define as $f_{\rm duty}\equiv \langle \ell^2 \rangle / \langle \ell\rangle^2$ \citep[see, e.g.,][]{Ciotti:01}, is in the limit $\ell_{\rm s.s.}\ll \ell_{\rm max}$ given by
\beq
f_{\rm duty} \sim \frac{2}{\ln(\ell_{\rm max}/\ell_{\rm s.s.})} \sim 2 \frac{\langle\ell\rangle}{\ell_{\rm max}} .
\eeq

The weak logarithmic dependence on the ratio of the maximum to the minimum luminosity is deceptive; while we generally expect a duty cycle in the range $f_{\rm duty}\sim 0.2-1$, the average accretion rate depends on the peak rate. We can attempt to generalize equation (\ref{eq:ell_bondi}) to model the peak rate by replacing $T_{\rm s}$ with $T_{\rm HII}$ (to mimic an initial absence of heating at the sonic radius) and by including the density enhancement factor $f_{\rm epi}$ to obtain
\beq
\label{eq:ell_bondi_infall}
\ell_{\rm max} \sim 0.03~ \frac{\epsilon_{-1} ~f_{\rm turb}~f_{\rm epi,1}~\Upsilon ~M_2 ~n_5 ~T_{{\rm HI},3.7}}{
T_{{\rm HII},4.7}^{5/2}} ,
\eeq
which would imply a duty cycle of $f_{\rm duty}\sim 0.6$ (for $f_{\rm epi}\sim10$) and an average accretion rate that is a factor of $\sim 9$ times higher than the steady state rate. 

The duty cycle estimated here differs from the one derived by \citet{Ricotti:08}, $f_{\rm duty,ROM}=(r_{\rm B}/r_{\rm ion})^{1/3}$, where for a stationary black hole $r_{\rm B}\equiv GM_{\rm BH}/c_{\rm s,HI}^2\sim 2.4\times10^{16} ~M_2 ~T_{{\rm HI},3.7}^{-1}\textrm{ cm}$ is the Bondi radius ignoring radiative feedback \citep{Bondi:44}.  This estimate assumes that the accretion is efficient only when $r_{\rm ion}\leq r_{\rm B}$.  In our picture, $r_{\rm B}$ does not have a unique physical meaning because the accretion flow structure is strongly modified by photoionization, radiative heating, and the radiation pressure; episodic accretion through the \ion{H}{2} region can proceed even when $r_{\rm ion}\gg r_{\rm B}$. 

The peak accretion rate $\ell_{\rm max}$ and the average rate $\langle \ell\rangle$ during episodic accretion can exceed the limits imposed by photoheating and radiation pressure because of the presence of the finite inertia of dense infalling shells and the lack of pressure equilibrium in the gas photoionized from the inner edge of the \ion{H}{2} region during luminosity maxima.
We are not able to determine with certainty whether the accretion from a weakly turbulent quasiuniform density medium will be episodic, in which case it will proceed at the rate given by equations (\ref{eq:ellbar_episodic}) and (\ref{eq:ell_bondi_infall}), or whether it will be steady at the rate given in equation (\ref{eq:ell_bondi}) subject to radiation pressure-suppression derived in \S~\ref{sec:prad_continuum}. We proceed to discuss the implications of inhomogeneity and turbulence in the environment of a seed black hole.

\section{Molecular Cooling and Clumpy Accretion}
\label{sec:clumpy}

\subsection{Accretion of Self-Shielding Clumps}
\label{sec:self_shielding_clumps}

As a protogalaxy grows, the baryonic inflow velocity into its center increases.  Cosmological hydrodynamic simulations show that baryons accreting along the filaments of the cosmic web can remain cold until they reach the central region of the protogalaxy.  Supersonic baryonic inflow, in protogalaxies with virial velocities $\gtrsim 10\textrm{ km s}^{-1}$, drives turbulence in the galaxy.  The resulting central turbulent Mach numbers measured in the simulations of $\sim 10^8~M_\odot$ cosmological halos are $\sim 3$ \citep{Wise:08,Greif:08}.  Supersonic turbulence is expected in view of the short cooling time $t_{\rm shock,cool}\sim 10\ n_5^{-1} \textrm{ yr}$ of the gas heated at the termination shocks of the cold inflows \citep[for somewhat higher velocity, $50\textrm{ km s}^{-1}$ shocks, e.g.,][see also \citealt{Gnat:08} for dependence on metallicity, which is weak for $Z\lesssim 0.1~Z_\odot$]{Shapiro:87,Kang:92}. If the shocked gas cools on a dynamical time to temperatures $\sim 10^4\textrm{ K}$, molecules form and the gas quickly cools further. Therefore, a fraction of the dense gas mass that is collecting at the center of the protogalaxy may reside at temperatures substantially below the $T_{\rm HI}\sim 5,000\textrm{ K}$ that we have somewhat arbitrarily taken as the temperature floor thus far.

Near an accreting seed black hole, molecule formation could be enhanced even in the absence of dust. Hard X-rays emitted by the black hole could maintain a high electron fraction in the neutral gas surrounding the \ion{H}{2} region, which would catalyze molecule formation. Molecules can also form if traces of metals and dust are present. The cooling to temperatures characteristic of the molecular phase leads to an increase of turbulent Mach numbers.  In supersonic turbulence, the local density exhibits intermittent strong fluctuations around the mean. Simulations of isothermal turbulence have shown that the density probability density function is a normal distribution in $\ln\rho$ with mean $\langle \ln\rho\rangle=-\frac{1}{2}\sigma^2$ and dispersion $\sigma^2 =\ln(1+b^2{\cal M}^2)$, where  ${\cal M}$ is the turbulent Mach number and $b\approx 0.26$ for unmagnetized turbulence \citep[e.g.,][and references therein]{Kritsuk:07}.  For example, for ${\cal M}=10$, $\sim 1\%$ of the volume contains densities in excess of the average density by a factor of $10$ or greater. Dense clumps can enter the ionized sphere and remain self-shielded from photoionization and photodissociation.  Since from equation (\ref{eq:rion}) the critical radius for photoionization is proportional to $n^{-2/3}$, a clump with an overdensity of $10$ can withstand photoionization to radii $\sim0.2\ r_{\rm ion}$.

The presence of neutral clumps in the ionized sphere may fundamentally alter the structure of the accretion flow if they are dense enough to survive photoionization and photoevaporation, which requires densities violating the condition in equation (\ref{eq:n_optically_thin}). 
Such dense clumps may form as a product of turbulent fragmentation and may be self-gravitating and on their way to turn into star-forming cores, provided that they are not tidally disrupted by and accreted onto the black hole \citep[e.g.,][]{Bonnell:08}. During accretion minima when $\ell\ll \ell_{\rm max}$, dense clumps may withstand photoionization and photoevaporation even at $r\lesssim r_{\rm s}$.  In this case, they could accrete directly into the small optically thick accretion disk around the black hole.  If the clumps occupy a substantial solid angle as seen from the black hole, the gas in their shadows recombines, cools, and accelerates toward the black hole.

We thus speculate that with the increase of turbulent Mach numbers and the progression of ever higher degree of clumping in the gas, the severity of the radiative feedback discussed in \S~\ref{sec:bondi} and \S~\ref{sec:timedep} decreases.  The feedback-limited accretion may proceed at only a small fraction of the rate corresponding to the Eddington limit until turbulent inhomogeneities reach the level at which the densest clumps are self-shielded at $\sim r_{\rm s}$ even at $\ell_{\rm max}\sim 1$.  Then, unless the gas supply is depleted by star formation and supernovae, the seed black hole may be able to grow efficiently and double its mass on the Salpeter time scale $t_{\rm Salp}\sim 5\times 10^7\ \epsilon_{-1}\textrm{ yr}$.  Central stellar velocity dispersions found in massive black hole-hosting stellar systems in the local universe are $> 30\textrm{ km s}^{-1}$ \citep[e.g.,][and references therein]{Barth:05}; if massive black holes are not found in systems with smaller dispersions, this may be interpreted as a hint that efficient accretion commences only when the baryonic gravitational potential well depth around the black hole exceeds a critical minimum value.

\subsection{Photodissociation Sphere around the Black Hole}
\label{sec:photodissociation}

The existence of dense clumps in the neighborhood of the \ion{H}{2} region may be contingent on the presence of molecules. The radiation from the black hole likely contains a photo-dissociating component below the Lyman edge.  However, for a central source with a hard spectrum, two photon emission from the $2^1{\rm S}\rightarrow 1^1{\rm S}$ transition in the recombination of \ion{He}{2} is a guaranteed source of H$_2$-dissociating photons.  Photons in the Lyman-Werner (LW) bands of hydrogen are produced at the rate $\dot N_{\rm LW}\sim \frac{1}{3} f_{\rm ion,He}L/E_{\rm He}$ \citep{Johnson:07}, where $f_{\rm ion,He}$ is the fraction of photon energy that goes into helium ionization in mostly neutral gas, and $E_{\rm He}=24.6\textrm{ eV}$. Some LW radiation may also be produced by the central source itself \citep{Kuhlen:05}. If the gas surrounding the black hole is dust-free and molecule synthesis is catalyzed primarily by H$^-$, one can estimate the dissociation depth $\Delta r_{\rm diss}$ in the neutral gas via
\beq
\label{eq:ndot_lw}
 \frac{4\pi}{3} (r_{\rm diss}^3-r_{\rm ion}^3) ~k_{\rm H_2} ~n_{\rm H^-} ~n_{\rm H} = \dot N_{\rm LW} ,
\eeq
where $r_{\rm diss} \equiv r_{\rm ion}+\Delta r_{\rm diss}$, $k_{\rm H_2}$ is the rate for the reaction ${\rm H}^-+{\rm H}\rightarrow {\rm H}_2+e^{-}$, and $n_{{\rm H}^-}$ is the equilibrium abundance of the H$^-$ ions. 
Adopting the temperature $T=1,000\textrm{ K}$ we have $k_{\rm H_2}\approx 1.2\times10^{-9}\textrm{ cm}^3\textrm{ s}^{-1}$ and $n_{\rm H^-}\sim 7\times10^{-7} n_e$ \citep[see, e.g.,][and references therein]{Oh:02}. In writing equation (\ref{eq:ndot_lw}), we have ignored the potential enhancement of H$_2$ formation rate in a supersonically turbulent gas, where molecule formation is particularly efficient in intermittent overdensities \citep[see, e.g.,][]{Pavlovski:02,Glover:07b}. 

The maximum ionization fraction $\chi_e \equiv n_e/n_{\rm H}$ allowed if H$_2$ is to be dissociated in a sphere of radius $2r_{\rm ion}$ is obtained by substituting $\Delta r_{\rm diss}=r_{\rm ion}$ in equation (\ref{eq:ndot_lw}) and solving for $\chi_e$, to obtain the condition
\beq
\label{eq:xe_max_dissoc}
\chi_e < 2.4\times 10^{-4}~  \frac{f_{\rm turb}^2~T_{{\rm HI},3.7}^2}{f_{\rm ion} ~ T_{{\rm HII},4.7}^{3}} ,
\eeq
where we have set $f_{\rm ion,He}=0.06$. If the SED of the black hole extends into the hard X-rays, an electron fraction violating the condition in equation (\ref{eq:xe_max_dissoc}) may be maintained in the neutral shell surrounding the \ion{H}{2} region.  This suggests that, even in an environment with primordial composition, the \ion{H}{2} region associated with a seed black hole may be surrounded by only a thin ($\Delta r_{\rm diss}\ll r_{\rm ion}$) photodissociation region where gas cooling and clumping is reduced \citep[see also][]{Johnson:07b}. Photodissociation may be even less important if the protogalaxy has been enriched with trace quantities of metals and dust \citep[e.g.,][]{Omukai:08}, as gas cooling can proceed even in the absence of molecular hydrogen in that case.  Also, if the accretion is episodic (\S~\ref{sec:timedep}), molecular gas can form during the quiescent periods in which central accretion is temporarily diminished or suspended by the radiative feedback.

We have ignored the photodissociation of H$^-$, which reduces the H$_2$ formation rate by a factor $S^{-1}$, where $S=1+\dot N_{\gamma,{\rm H}^-}\sigma_{{\rm H}^-}/4\pi r^2 k_{{\rm H}_2} n_{\rm H}$, $\dot N_{\gamma,{\rm H}^-}$ is the total dissociating photon number luminosity, and $\sigma_{{\rm H}^-}$ is the average photodissociation cross section \citep[see, e.g.,][and references therein]{Glover:06,Glover:07a,Yoshida:07,Chuzhoy:07}.  In our estimate of $\dot N_{\gamma,{\rm H}^-}$, we will ignore the multiplying effect of Ly$\alpha$ trapping (\S~\ref{sec:prad_lyalpha}) on the H$^-$-photodissociating photon number density; a more accurate approach, which would be too involved to include here, would be to solve for Ly$\alpha$ resonance line and X-ray continuum transfer in the  H$^-$ photodissociation region self-consistently. 
\citet{Chuzhoy:07} estimate that for reprocessed ionizing radiation, the average cross section per recombination photon times the average number of photons per recombination varies in the range $\langle \sigma_{{\rm H}^-}\rangle\sim (1.6-3.4)\times10^{-17}\textrm{ cm}^2$. Setting $\dot N_{\gamma,{\rm H}^-}\sim \dot N_{\rm rec} $, where $\dot N_{\rm rec} \sim (4\pi/3) \alpha_B(T_{\rm HII}) n_{{\rm H}^+}n_e$ is the hydrogen recombination rate inside the \ion{H}{2} region, we find
\beq
S\sim 1 +  0.43~ \frac{\ell^{1/3} ~f_{\rm ion}^{1/3}~ f_{\rm turb}^{4/3}~ M_2^{1/3} ~n_5^{1/3}~ T_{{\rm HI},3.7}^{4/3}}{ T_{{\rm HII},4.7}^{2}} \left(\frac{r}{r_{\rm ion}}\right)^{-2},
\eeq
which shows that H$_2$ formation suppression  is marginally important in the case of an accretion from a high density cloud.  The suppression widens the photodissociation region surrounding the \ion{H}{2} region.  

Having raised the possibility of a multiplying effect of Ly$\alpha$ trapping on the photodissociation of H$^-$, we remark that the destruction of Ly$\alpha$ photons by H$^-$, which is a processes that have ignored in \S~\ref{sec:prad_lyalpha}, may have to be considered in a self-consistent study of Ly$\alpha$ transfer in a gas with primordial composition. Because of significant uncertainties in the ionization fraction and the H$^-$ abundance in the neutral shell, which both depend on the SED are not calculated accurately in our simplified model, we attempt only a crude estimate.  According to equation (7) of \citet{Hansen:06}, Ly$\alpha$ photons scattering through a shell of column depth $N$ are destroyed before escaping if 
\beq
N> 6.7\times10^{19}~\frac{T_{{\rm HI},3.7}^{1/4}}{\chi_{{\rm H}^-}^{3/4}}~ \left(\frac{\sigma_{{\rm Ly}\alpha,{\rm H}^-}}{10^{-21}\textrm{ cm}^2}\right)^{-3/4}\textrm{ cm}^{-2},
\eeq
where $\sigma_{{\rm Ly}\alpha,{\rm H}^-}\sim 4.4\times10^{-18}\textrm{ cm}^{-2}$ is the H$^-$ photodissociation cross section by Ly$\alpha$, and $\chi_{{\rm H}^-}$ is the H$^-$ abundance.  Setting $N\sim n_{\rm H} \Delta r $ we find that Ly$\alpha$ photons are destroyed before escaping the shell of width $\Delta r$ when the ionization fraction in the shell is
\bea
\chi_e &>& 0.0024 ~ \frac{S~f_{\rm turb}^{8/9} 
 ~ T_{{\rm HI},3.7}^{11/9} ~
}{ \ell^{4/9} ~f_{\rm ion}^{4/9} ~n_5^{4/9} ~ M_2^{4/9}~T_{{\rm HII},4.7}^{4/3} } \left(\frac{\Delta r}{r_{\rm ion}}\right)^{-4/3} .
\eea
This crude estimate suggests that for accretion in a dense, dust-free medium with a central X-ray source that maintains a high ionization fraction in the neutral gas, Ly$\alpha$ photons may be destroyed by H$^-$ absorption before penetrating the shell.  The issue warrants a self-consistent treatment of resonance line scattering and continuum radiation transfer coupled to the chemistry of the \ion{H}{2} region and the surrounding neutral shell.  

\section{Conclusions}
\label{sec:conclusions}

We model quasiradial accretion onto seed massive black holes in metal and dust poor protogalaxies with the goal of evaluating the common assumption that the $M_{\rm BH}\gtrsim 100~M_\odot$ seed black holes accrete at the Bondi accretion rate moderated by the Eddington limit. After considering radiative feedback effects in the neighborhood of an accreting black hole, we are able to derive the following conclusions:

The local thermal and statistical equilibrium temperature of a photoionized gas is a strong function of the metallicity of the gas at radii from the black hole where gas becomes captured by the black hole.  
The photoionized gas outside the sonic radius is normally in equilibrium, but, particularly for metallicities $Z\lesssim 0.1\ Z_\odot$, the gas passing through the sonic radius may not be in equilibrium.  For radiative efficiencies $\epsilon\lesssim 0.1$, the gas will  not complete thermal runaway that would heat it to the Compton temperature prior to crossing the sonic radius

Due to the radiative heating of the gas, the sonic radius is much smaller than the Bondi radius evaluated in terms of the temperature of the protogalactic medium far from the black hole; the corresponding gasdynamical accretion rate is thus reduced to a small fraction of the Bondi value.  The gas temperature at the sonic radius is also a strong function of the radiative efficiency of the accretion near the event horizon of the black hole. 

The black hole is surrounded by an \ion{H}{2} region, which is further surrounded by a photodissociation region if the ambient protogalactic medium contains molecules. The radius of the \ion{H}{2} region in dense ($n\gtrsim 10^5\textrm{ cm}^{-3}$) clouds, that must be present to support near-Eddington accretion, is small ($r_{\rm ion}\sim 1\textrm{ pc}$) compared to the size of the protogalaxy.  The sonic radius is smaller by orders of magnitude than the radius of the \ion{H}{2} region.  This vast separation of scales renders self-consistent simulations of accretion onto seed black holes from a protogalactic environment particularly challenging, because to appropriately capture the radiative effects, the simulations must at least resolve the sonic radius.

Photoionization radiation pressure from the continuum radiation produced near the event horizon may further diminish the accretion rate. Compared to the gravitational acceleration, the acceleration due to radiation pressure is the strongest in the bulk of the \ion{H}{2} region, i.e., not too close to the black hole where the equilibrium neutral fraction is negligible.  We have derived a simple criterion for accretion rate suppression by the photoionization radiation pressure.

A significant fraction of the luminosity of the black hole is converted into Ly$\alpha$  resonance line radiation; resonance line scattering keeps this radiation trapped in the neighborhood of the \ion{H}{2} region.  Radiation pressure from the trapped radiation can exceed the external, confining gas pressure.  We do not self-consistently solve for a stationary accretion flow in the presence of resonance line radiation pressure, but our estimates do suggests that the resonance line pressure could be detrimental to stationary accretion at high accretion rates.

As an alternative to strictly stationary accretion, we outline a model for episodic accretion in which short episodes of sudden, rapid accretion alternate with longer periods of accretion reduced by photoheating and radiation pressure.  In this model, the outflow and central rarefaction induced by photoheating and radiation pressure acceleration in the bulk of the \ion{H}{2} region drives down the central accretion rate rapidly. The time-average accretion rate is then set by the peak accretion rate, divided by the logarithm of the number of $e$-foldings in the outflow-induced accretion rate decay.  Since the peak accretion rate need not be limited by the thermodynamic and kinematic constraints that apply to steady-state solutions, the episodic accretion rate may exceed the steady-state rate by a large factor.

Our idealized treatment here already indicates how extremely complex the
self-consistent accretion problem is. To more fully explore the
time-dependent accretion flow onto a seed black hole, numerical simulations
are clearly needed. Of particular importance is to study the effect of angular momentum,
the impact of turbulence, and the possible emergence of a multi-phase medium
in the infalling gas. We will report on the results from such simulations and on a comparison with the analysis presented in this work elsewhere \citep[][Couch et al., in preparation]{Milosavljevic:08}.  If our result that the initial accretion onto stellar seed black holes is greatly reduced in the presence of radiative feedback holds up even in fully three dimensional radiation-hydrodynamical simulations, the need for massive black hole formation by radiatively-inefficient accretion or by direct collapse of massive primordial gas clouds might be significantly increased \citep[e.g.,][]{Bromm:03,Begelman:06,Begelman:08,Djorgovski:08}. 

\acknowledgements

We thank an anonymous referee for detailed and very helpful comments and suggestions.
V.~B. and M.~M. acknowledge support from NSF grant AST-0708795. 
S.~P.~O. acknowledges support from NASA grant NNG06GH95G.

\end{document}